\documentclass[prd,twocolumn,floatfix,amsmath,nofootinbib,amssymb,floatfix]{revtex4}
\usepackage{graphicx,color,dcolumn,booktabs,bm}
\usepackage{longtable,lscape}
\usepackage{pdfpages}
\usepackage{txfonts}
\usepackage{overpic}
\usepackage{amssymb}
\usepackage{makecell}
\usepackage{indentfirst}
\usepackage{feynmf}   
\usepackage{slashed}  
\usepackage{cases}
\usepackage{color}
\usepackage{multirow}
\usepackage{threeparttable}
\usepackage{epstopdf}
\usepackage{enumerate}
\usepackage{subfigure}
\usepackage{diagbox}
\usepackage{graphicx,color,dcolumn,booktabs,bm}
\usepackage{mathrsfs}
\usepackage{cancel}
\usepackage{float}
\usepackage[colorlinks,
            citecolor=blue,
            anchorcolor=red,
            menucolor=red,
            linkcolor=red,
            filecolor=red,
            runcolor=red,
            urlcolor=blue,
            frenchlinks=red]{hyperref}

\begin{document}
\title{Heavy-quark dominance and fine structure of excited heavy baryons $\Sigma_{Q}$, $\Xi '_{Q}$ and $\Omega_{Q}$}
\author{Zhen-Yu Li$^{1,4}$}
\email{zhenyvli@163.com }
\author{Guo-Liang Yu$^{2}$}
\email{yuguoliang2011@163.com }
\author{Zhi-Gang Wang$^{2}$ }
\email{zgwang@aliyun.com }
\author{Jian-Zhong Gu$^{3}$ }
\email{gujianzhong2000@aliyun.com }

\affiliation{$^1$ School of Physics and Electronic Science, Guizhou Education University, Guiyang 550018,
China\\$^2$ Department of Mathematics and Physics, North China Electric Power University, Baoding 071003,
China\\$^3$ China Institute of Atomic Energy, Beijing 102413,China\\$^4$ Guangxi Key Lab of Nuclear Physics and Technology, Guangxi Normal University, Guilin 541006, China}
\date{\today }

\begin{abstract}
In the framework of the relativized quark model, the calculation of spin-orbit interactions is improved by considering the contribution from the light-quark cluster in a singly heavy baryon. It modifies the energy level splitting of the orbital excitation significantly and causes the emergence of fine structures for $\Sigma_{Q}$, $\Xi '_{Q}$ and $\Omega_{Q}$ baryons. Based on this improvement, we systematically analyze the fine structures and retest the heavy quark dominance mechanism. This mechanism is found to be violated in the $1P$-wave states of the $\Sigma_{c}$, $\Xi '_{c}$ and $\Omega_{c}$ baryons although it remains effective overall, which may help to understand the nature of the heavy quarks and strong interactions. With the predicted fine structures, we make the precise assignments of those observed heavy baryons which once could not be accurately explained due to their close mass values. The method used in this work is instructive and applicable for the study of more complex exotic hadrons, such as the heavy tetraquarks and pentaquarks.

\end{abstract}

\maketitle

\section{Introduction}\label{sec1}

The heavy baryon spectroscopy has played an important role in the development of Quantum Chromodynamics (QCD)~\cite{F101}. It provides an ideal place to understand the properties of heavy quarks, such as the heavy quark symmetry, chiral dynamics, feature of strong interactions, and relevant models inspired by QCD. So far, a large number of singly heavy baryons have been observed in experiment, which has advanced the study of spectroscopy significantly~\cite{F201,F2021,F205,F206,F207,F208,F210,F209,F211,F212,F213,F4401,F4402,F4403,F203,F204,F214}.
In the new summary tables of the particle data group (PDG), about 70 singly heavy baryons have been collected~\cite{F201}. Some of them, however, their spins and parities ($J^{P}$) have not yet been identified, especially for the $\Sigma_{c(b)}$, $\Xi '_{c(b)}$ and $\Omega_{c(b)}$ families.

In fact, as early as 2004, $\Sigma_{c}(2800)^{0,+,++}$ were observed experimentally by Belle collaboration~\cite{F210}, and later confirmed by BaBar collaboration~\cite{F209}. 20 years later, the $J^{P}$ quantum numbers of $\Sigma_{c}(2800)^{0,+,++}$ have yet to be determined. As a possible member of the $\Sigma_{c}$ family, the discovery of $\Sigma_{c}(2846)^{0}$ was announced by BaBar collaboration in 2008~\cite{F209}, nevertheless, which has not been further confirmed by theory and experiment. Similarly, $\Sigma_{b}(6097)^{+,-}$ ~\cite{F211} and $\Xi_{b}(6227)^{0,-}$~\cite{F212,F213} were observed in 2018, respectively. And their $J^{P}$ values have not yet been fixed.

On the other hand, many excited heavy baryons were observed in groups. In 2017, LHCb collaboration reported the observation of five narrow $\Omega_{c}^{0}$ states~\cite{F4401}: $\Omega_{c}^{0}(3000)$, $\Omega_{c}^{0}(3050)$, $\Omega_{c}^{0}(3065)$, $\Omega_{c}^{0}(3090)$ and $\Omega_{c}^{0}(3120)$. Later, Belle collaboration unambiguously confirmed four of these states, but no signal was found for the $\Omega_{c}^{0}(3120)$~\cite{F4402}. In 2020, the first observation of four excited $\Omega_{b}^{-}$ states was announced by LHCb collaboration~\cite{F4403}. They are $\Omega_{b}(6316)^{-}$, $\Omega_{b}(6330)^{-}$, $\Omega_{b}(6340)^{-}$ and $\Omega_{b}(6350)^{-}$. With the discovery of these excited $\Omega_{c(b)}$ states, $\Xi_{c}(2923)^{0}$, $\Xi_{c}(2939)^{0}$, $\Xi_{c}(2964)^{0}$~\cite{F203} and $\Xi_{c}(2930)^{+}$~\cite{F204} have been observed, respectively. Most of the above baryons have been collected in the new heavy baryon list of PDG~\cite{F201}, but their $J^{P}$ values are still unconfirmed.

The studies of these excited baryons have been performed theoretically with a vast variety of approaches, such as non-relativistic quark models~\cite{F403,Fb05,Fb08}, relativized quark models~\cite{F401,F402,Fp003}, hypercentral quark models~\cite{F3171,F3311}, chiral quark model~\cite{F322,F323}, constituent quark model~\cite{Fp005,F3221,F3241,F324}, harmonic oscillator based models~\cite{Fb003,F334}, QCD sum rules~\cite{F315,F320,F333,Fb002,Fb06,Fb0021,Fb061}, quark-diquark pictures~\cite{F405,F407,Fp006,F4071}, effective field theories~\cite{F303,F304,F3231}, Lattice QCD~\cite{Fb04}, and $^{3}P_{0}$ models~\cite{Fb01,Fb011,Fp007,Fb07}. Their predicted $J^{P}$ values are collected in Table~\ref{ta01}. As is shown in Table~\ref{ta01}, different $J^{P}$ values of these excited baryons are predicted by different models or theories, which indicates that the theoretical studies have not yet reached a consensus.

In the aspect of theory, the heavy-quark in a singly heavy baryon is decoupled from the two light-quarks in the heavy quark limit. With the requirement of the flavor $SU(3)$ subgroups for the light-quarks, the baryons belong to either a sextet ($\mathbf{6}_{F}$) of the flavor symmetric states ($\Sigma_{Q}$, $\Xi'_{Q}$ and $\Omega_{Q}$) or an anti-triplet ($\mathbf{\bar{3}}_{F}$) of the flavor antisymmetric states ($\Lambda_{Q}$ and $\Xi_{Q}$)~\cite{F403}. Here $Q$ denotes $c$ quark or $b$ quark. The theoretical research shows that the above mentioned excited baryons most likely belong to the $1P$-wave states. However, a further confirmation of the $J^{P}$ values of these baryons is very difficult.

The theoretical analysis of these particles is challenging for three primary reasons. The first reason is that the masses of these excited baryons in each family (such as $\Xi_{c}^{'}$, $\Omega_{c}$ and $\Omega_{b}$ families) are very close to each other. It requires more precise theoretical calculations. The second one is the lack of more systematic theoretical calculations. While the systematic theoretical work is essential to enhance the reliability of theoretical predictions. Additionally, for these excited baryons, such close mass values in each group indicate a fine structure in their orbital excitation spectra. However, the fine structure in the heavy baryon spectroscopy is still an unsolved problem.
So, the reasonable explanation for these excited baryons is currently an interesting challenge for theoretical research, and also an opportunity to understand the fine structure of the heavy baryon spectra. And a more systematic theoretical analysis is requisite.

Inspired by some related theoretical works~\cite{F401,F402,F403,F405,F407,F410}, we studied the spectra of singly and doubly heavy baryons systematically in the framework of the relativized quark model. The results show that most of the available experimental data can be well described with a uniform set of parameters~\cite{F502,F503,F504,F505,F506}. This set of parameters used in calculations is consistent with that originally appeared in the model developed by Godfrey and Isgur (the GI model~\cite{F401}), apart from the two parameters related to the linear confinement potential~\cite{F502,F503}. The successful description of the experimental data reflects the systematic and reliable nature of the relativized quark model.

Based on the above systematic works, we proposed the heavy-quark dominance (HQD) mechanism, which gives a physical explanation of the orbital excitation~\cite{F501}. The HQD mechanism determines the overall structure of excitation spectra in singly and doubly heavy baryons. The implementation of this mechanism may provide a pathway for a further exploration in the spectroscopy of heavy baryons. However, our current research on the heavy baryon spectroscopy still faces two problems below.

\emph{(1) HQD mechanism breaking for the baryons $\Sigma_{c}$, $\Xi '_{c}$ and $\Omega_{c}$.}
The HQD mechanism means that the orbital excitation mode with lower energy levels is always dominated by the heavy quark(s) in the singly (doubly) heavy baryon, if the mass of the heavy quark(s) is large enough. This mechanism is well represented in the bottom ($b$) baryons. However, our calculation shows that the HQD mechanism is not obvious for the orbital excitation of the singly heavy baryons $\Sigma_{c}$, $\Xi '_{c}$ and $\Omega_{c}$~\cite{F501}. The reason is that charm ($c$) quark is not so heavy. Then, the HQD mechanism could be broken for these charm baryons, which will directly affect the structure of the excitation spectrum.

\emph{(2) Fine structure of the heavy baryons $\Sigma_{c(b)}$, $\Xi '_{c(b)}$ and $\Omega_{c(b)}$.}
For the orbital excitation spectra of the $\Sigma_{c(b)}$, $\Xi '_{c(b)}$ and $\Omega_{c(b)}$ families, the fine structure predicted by our calculations has serious energy level degeneracy and does not fit well to these data~\cite{F501}. The reason might be that the form of spin-orbit interactions which we used is too simple.
So, the fine structure of the excited baryon spectrum may be explained by improving the calculation of spin-orbit interactions. Actually, the spin-orbit force plays an important role at different levels of matter~\cite{F906,F905,F904,F903,F902}. In the case of baryon systems, the mechanism of spin-orbit interactions is not fully understood, and has always puzzled the theorists~\cite{F306,Fp001,Fp002}. A deeper understanding of the spin-orbit interactions is also a topic of current research in hadron physics.

In this work, we will try to improve the calculation of the spin-orbit interactions, so as to provide the solution to these two problems and give a more reliable theoretical analysis of these excited heavy baryons.
The remainder of this paper is organized as follows. In
Sec.~\ref{sec2}, the theoretical methods used in this work are introduced, including the Hamiltonian of the relativized quark model, the structure of the wave function and the improved calculation of the Hamiltonian eigenvalues including the contribution from the light-quark cluster to the spin-orbit interactions. The fine structure of the excitation spectra, the retest of the HQD mechanism and the assignments of the observed baryons are analyzed in Sec.~\ref{sec3}. And Sec.~\ref{sec4} is reserved for the conclusions.

\begin{table*}[htbp]
\begin{ruledtabular}\caption{Some predicted $J^{P}$ values for the $1P$-wave excited baryons of the $\Sigma_{c(b)}$, $\Xi^{'}_{c(b)}$ and $\Omega_{c(b)}$ families. The experimental data are taken by their isospin averages. }
\label{ta01}
\begin{tabular}{c c c c c c c c c }
Baryon/$M_{exp.}$(MeV)   & \multicolumn{7}{c}{Predicted $J^{P}$ values}    \\ \cline{1-1} \cline{2-9}  \\
$\Sigma_{c}(2800)^{++,+,0}$/$\sim$2800~\cite{F201} & 1/2$^{-}$~\cite{F403}  &  3/2$^{-}$~\cite{F403} &  5/2$^{-}$~\cite{F403} &3/2$^{-}$~\cite{Fp003} & 3/2$^{-}$~\cite{Fp003}  &  5/2$^{-}$~\cite{Fp003} & 1/2$^{-}$~\cite{F405}  & 3/2$^{-}$~\cite{F405} \\
$\Sigma_{c}(2846)^{0}$/$\sim$2846~\cite{F209} & 1/2$^{+}$~\cite{F407}   \\\\
$\Sigma_{b}(6097)^{+,-}$/$\sim$6097~\cite{F201}& 1/2$^{-}$~\cite{Fp003}  & 3/2$^{-}$~\cite{Fp003} &3/2$^{-}$~\cite{F324} & 5/2$^{-}$~\cite{Fp006} &3/2$^{-}$~\cite{Fp006}  & 1/2$^{-}$~\cite{Fb07} \\\\
$\Xi_{c}(2923)^{0}$/$\sim$2923~\cite{F203} & 1/2$^{-}$~\cite{Fp003}  & 3/2$^{-}$~\cite{Fp003} & 1/2$^{-}$~\cite{Fp007} & 1/2$^{-}$~\cite{Fb061} & 3/2$^{-}$~\cite{Fb011}  \\
$\Xi_{c}(2939)^{0}$/$\sim$2939~\cite{F203}  & 3/2$^{-}$~\cite{Fp007}  & 3/2$^{-}$~\cite{Fb061} & 5/2$^{-}$~\cite{Fb011} \\
$\Xi_{c}(2964)^{0}$/$\sim$2965~\cite{F203} & 3/2$^{-}$~\cite{Fp007}  & 3/2$^{-}$~\cite{Fb061} & 1/2$^{+}$~\cite{Fb011}  \\
$\Xi_{c}(2930)^{+}$/$\sim$2942~\cite{F204} & 1/2$^{-}$~\cite{Fp003}  & 3/2$^{-}$~\cite{Fp003} &  1/2$^{-}$~\cite{F3171} &3/2$^{-}$~\cite{Fp007} \\\\
$\Xi_{b}(6227)^{0,-}$/$\sim$6227~\cite{F201} & 1/2$^{+}$~\cite{Fp003}  & 3/2$^{-}$~\cite{Fp003} & 3/2$^{-}$~\cite{F3311} & 5/2$^{-}$~\cite{F324} & 5/2$^{-}$~\cite{Fp007}  & 5/2$^{-}$~\cite{Fb07} \\\\
$\Omega_{c}(3000)^{0}$/$\sim$3000~\cite{F201} & 1/2$^{-}$~\cite{Fb05}  & 1/2$^{-}$~\cite{Fb08} & 1/2$^{-}$~\cite{Fp003} &  1/2$^{-}$~\cite{F3241}  & 1/2$^{-}$~\cite{Fb06} & 1/2$^{-}$~\cite{Fb04} & 1/2$^{+}$~\cite{Fb01}  &3/2$^{+}$~\cite{Fb01}\\
$\Omega_{c}(3050)^{0}$/$\sim$3050~\cite{F201} & 1/2$^{-}$~\cite{Fb05}  & 1/2$^{-}$~\cite{Fb08} & 3/2$^{-}$~\cite{Fp003} & 3/2$^{-}$~\cite{F3241} & 3/2$^{-}$~\cite{Fb06}  & 1/2$^{-}$~\cite{Fb002} & 1/2$^{-}$~\cite{F3231} & 1/2$^{-}$~\cite{Fb04} \\ & 5/2$^{+}$~\cite{Fb01} & 7/2$^{+}$~\cite{Fb01}\\
 $\Omega_{c}(3065)^{0}$/$\sim$3065~\cite{F201}  & 3/2$^{-}$~\cite{Fb05}  & 3/2$^{-}$~\cite{Fb08} & 3/2$^{-}$~\cite{Fp003} & 3/2$^{-}$~\cite{F3241} & 1/2$^{+}$~\cite{Fb06} & 3/2$^{-}$~\cite{Fb002} & 3/2$^{-}$~\cite{Fb04} & 3/2$^{-}$~\cite{Fb01}\\
$\Omega_{c}(3090)^{0}$/$\sim$3090~\cite{F201} &3/2$^{-}$~\cite{Fb05}  & 3/2$^{-}$~\cite{Fb08} & 5/2$^{-}$~\cite{Fp003} & 5/2$^{-}$~\cite{F3241}&3/2$^{-}$~\cite{Fb002}  & 1/2$^{-}$~\cite{F3231} & 3/2$^{-}$~\cite{Fb04} & 5/2$^{-}$~\cite{Fb01} \\
$\Omega_{c}(3120)^{0}$/$\sim$3119~\cite{F201} & 5/2$^{-}$~\cite{Fb05}  & 5/2$^{-}$~\cite{Fb08} & 3/2$^{-}$~\cite{Fp003} & 3/2$^{+}$~\cite{Fb06} & 5/2$^{-}$~\cite{Fb002} & 5/2$^{+}$~\cite{Fb01} & 7/2$^{+}$~\cite{Fb01}  \\\\
$\Omega_{b}(6316)^{-}$/$\sim$6315~\cite{F201} & 1/2$^{-}$~\cite{Fb08}  & 1/2$^{-}$~\cite{Fp003} & 3/2$^{-}$~\cite{F3311} & 1/2$^{-}$~\cite{Fb07} & 3/2$^{-}$~\cite{Fb0021}  \\
$\Omega_{b}(6330)^{-}$/$\sim$6330~\cite{F201} & 3/2$^{-}$~\cite{Fb08}  & 3/2$^{-}$~\cite{Fp003} & 1/2$^{-}$~\cite{F3311} & 1/2$^{-}$~\cite{Fb07} & 1/2$^{-}$~\cite{Fb0021}  \\
$\Omega_{b}(6340)^{-}$/$\sim$6340~\cite{F201} & 1/2$^{-}$~\cite{Fb08}  & 3/2$^{-}$~\cite{Fp003} & 3/2$^{-}$~\cite{F3311} & 3/2$^{-}$~\cite{Fb07} & 5/2$^{-}$~\cite{Fb0021} \\
$\Omega_{b}(6350)^{-}$/$\sim$6350~\cite{F201} & 3/2$^{-}$~\cite{Fb08}  & 5/2$^{-}$~\cite{Fp003} & 3/2$^{-}$~\cite{F3311} & 3/2$^{-}$~\cite{Fb07} & 3/2$^{-}$~\cite{Fb0021} \\
\end{tabular}
\end{ruledtabular}
\end{table*}

\section{Theoretical methods used in this work}\label{sec2}
The relativized quark model was developed by Godfrey and Isgur to analyze meson spectra in 1985~\cite{F401}. Later, in 1986, Capstick and Isgur extended this model. They insisted on using the method of studying light baryons and systematically studied the mass spectra of light and heavy baryons under a unified framework~\cite{F402}. Very recently, Weng, Deng and Zhu followed this path and further advanced the study of baryon spectra~\cite{Fp003}.

Our works use the same model, but in a different way~\cite{F502,F503}. The spin-orbit terms used in our works are the quasi-two-body spin-orbit forces~\cite{Fp002}, which will be discussed in detail below.
\subsection{Hamiltonian of the relativized quark model}\label{sec2.1}
In the relativized quark model, the Hamiltonian for a three-quark system reads,
\begin{eqnarray}\label{e1}
\notag
H &&=H_{0}+\tilde{H}^{conf}+\tilde{H}^{hyp}+\tilde{H}^{so}\\
&&=\sum_{i=1}^{3}\sqrt{p_{i}^{2}+m_{i}^{2}}+\sum _{i<j}(\tilde{H}^{conf}_{ij}+\tilde{H}^{hyp}_{ij}+\tilde{H}^{so}_{ij}),
\end{eqnarray}
where the interaction terms $\tilde{H}^{conf}_{ij}$, $\tilde{H}^{hyp}_{ij}$ and $\tilde{H}^{so}_{ij}$ are the confinement, hyperfine and spin-orbit interactions, respectively.
The interactions are decomposed as follows~\cite{F401,F402}:
 \begin{eqnarray}
 \begin{aligned}
&\tilde{H}^{conf}_{ij}=G'_{ij}(r)+\tilde{S}_{ij}(r), \\
&\tilde{H}^{hyp}_{ij}=\tilde{H}^{tensor}_{ij}+\tilde{H}^{c}_{ij},\\
&\tilde{H}^{so}_{ij}=\tilde{H}^{so(v)}_{ij}+\tilde{H}^{so(s)}_{ij},
\end{aligned}
\end{eqnarray}
with
 \begin{eqnarray}
 &\tilde{H}^{tensor}_{ij}=-\frac{\textbf{s}_{i}\cdot\textbf{r}_{ij}\textbf{s}_{j}\cdot\textbf{r}_{ij}/r^{2}_{ij}-\frac{1}{3}\textbf{s}_{i}\cdot\textbf{s}_{j}}{m_{i}m_{j}}
\times(\frac{\partial^{2}}{\partial{r^{2}_{ij}}}-\frac{1}{r_{ij}}\frac{\partial}{\partial{r_{ij}}})\tilde{G}^{t}_{ij}, \\
&\tilde{H}^{c}_{ij}=\frac{2\textbf{s}_{i}\cdot\textbf{s}_{j}}{3m_{i}m_{j}}\nabla^{2}\tilde{G}^{c}_{ij},\\
\label{e5}
&\tilde{H}^{so(v)}_{ij}=\frac{\textbf{s}_{i}\cdot\textbf{L}_{(ij)i}}{2m^{2}_{i}r_{ij}}\frac{\partial\tilde{G}^{so(v)}_{ii}}{\partial{r_{ij}}}+
\frac{\textbf{s}_{j}\cdot\textbf{L}_{(ij)j}}{2m^{2}_{j}r_{ij}}\frac{\partial\tilde{G}^{so(v)}_{jj}}{\partial{r_{ij}}}+
\frac{(\textbf{s}_{i}\cdot\textbf{L}_{(ij)j}+\textbf{s}_{j}\cdot\textbf{L}_{(ij)i})}{m_{i}m_{j}r_{ij}}\frac{\partial\tilde{G}^{so(v)}_{ij}}{\partial{r_{ij}}}, \\
\label{e6}
&\tilde{H}^{so(s)}_{ij}=-\frac{\textbf{s}_{i}\cdot\textbf{L}_{(ij)i}}{2m^{2}_{i}r_{ij}}\frac{\partial\tilde{S}^{so(s)}_{ii}}{\partial{r_{ij}}}-
\frac{\textbf{s}_{j}\cdot\textbf{L}_{(ij)j}}{2m^{2}_{j}r_{ij}}\frac{\partial\tilde{S}^{so(s)}_{jj}}{\partial{r_{ij}}}.
\end{eqnarray}
Here, the following conventions are used, i.e., $\textbf{L}_{(ij)i}=\mathbf{r}_{ij}\times\mathbf{p}_{i}$ and $\textbf{L}_{(ij)j}=-\mathbf{r}_{ij}\times\mathbf{p}_{j}$. In the formulas above, $G'_{ij}$, $\tilde{G}^{t}_{ij}$, $\tilde{G}^{c}_{ij}$, $\tilde{G}^{so(v)}_{ij}$ and $\tilde{S}^{so(s)}_{ii}$ should be modified with the momentum-dependent factors as follows,
\begin{eqnarray}
\begin{aligned}\label{e7}
&G'_{ij}=(1+\frac{p^{2}_{ij}}{E_{i}E_{j}})^{\frac{1}{2}}\tilde{G}_{ij}(r_{ij})(1+\frac{p^{2}_{ij}}{E_{i}E_{j}})^{\frac{1}{2}}, \\
&\tilde{G}^{t}_{ij}=(\frac{m_{i}m_{j}}{E_{i}E_{j}})^{\frac{1}{2}+\epsilon_{t}}\tilde{G}_{ij}(r_{ij})(\frac{m_{i}m_{j}}{E_{i}E_{j}})^{\frac{1}{2}+\epsilon_{t}},\\
&\tilde{G}^{c}_{ij}=(\frac{m_{i}m_{j}}{E_{i}E_{j}})^{\frac{1}{2}+\epsilon_{c}}\tilde{G}_{ij}(r_{ij})(\frac{m_{i}m_{j}}{E_{i}E_{j}})^{\frac{1}{2}+\epsilon_{c}},\\
&\tilde{G}^{so(v)}_{ij}=(\frac{m_{i}m_{j}}{E_{i}E_{j}})^{\frac{1}{2}+\epsilon_{so(v)}}\tilde{G}_{ij}(r_{ij})(\frac{m_{i}m_{j}}{E_{i}E_{j}})^{\frac{1}{2}+\epsilon_{so(v)}}, \\
&\tilde{S}^{so(s)}_{ii}=(\frac{m_{i}m_{i}}{E_{i}E_{i}})^{\frac{1}{2}+\epsilon_{so(s)}}\tilde{S}_{ij}(r_{ij})(\frac{m_{i}m_{i}}{E_{i}E_{i}})^{\frac{1}{2}+\epsilon_{so(s)}},
\end{aligned}
\end{eqnarray}
where $E_{i}=\sqrt{m^{2}_{i}+p^{2}_{ij}}$ is the relativistic kinetic energy, and $p_{ij}$ is the momentum magnitude of either of the
quarks in the CM frame of the $ij$ quark subsystem~\cite{F402,F601}.

$\tilde{G}_{ij}(r_{ij})$ and $\tilde{S}_{ij}(r_{ij})$ are obtained by the smearing transformations of the one-gluon exchange potential $G(r)=-\frac{4\alpha_{s}(r)}{3r}$ and
linear confinement potential $S(r)=\tilde{b}r+\tilde{c}$, respectively,
\begin{eqnarray}\label{e8}
&\tilde{G}_{ij}(r_{ij})=\textbf{F}_{i}\cdot\textbf{F}_{j} \sum^{3}_{k=1}\frac{2\alpha_{k}}{\sqrt{\pi}r_{ij}}\int^{\tau_{kij}r_{ij}}_{0}e^{-x^{2}}\mathrm{d}x,
\end{eqnarray}
\begin{eqnarray}\label{e9}
\notag
\tilde{S}_{ij}(r_{ij}) &&=-\frac{3}{4}\textbf{F}_{i}\cdot\textbf{F}_{j} \{\tilde{b}r_{ij}[\frac{e^{-\sigma^{2}_{ij}r^{2}_{ij}}}{\sqrt{\pi}\sigma_{ij} r_{ij}}\\
&&+(1+\frac{1}{2\sigma^{2}_{ij}r^{2}_{ij}})\frac{2}{\sqrt{\pi}}\int^{\sigma_{ij}r_{ij}}_{0}e^{-x^{2}}\mathrm{d}x]+\tilde{c} \},
\end{eqnarray}
with
\begin{eqnarray}
\begin{aligned}
&\tau_{kij}=\frac{1}{\sqrt{\frac{1}{\sigma^{2}_{ij}}+\frac{1}{\gamma^{2}_{k}}}}, \\
&\sigma_{ij}=\sqrt{s^{2}_{0}(\frac{2m_{i}m_{j}}{m_{i}+m_{j}})^{2}+\sigma^{2}_{0}[\frac{1}{2}(\frac{4m_{i}m_{j}}{(m_{i}+m_{j})^{2}})^{4}+\frac{1}{2}]}.
\end{aligned}
\end{eqnarray}
Here $\alpha_{k}$ and $\gamma_{k}$ are constants. $\textbf{F}_{i}\cdot\textbf{F}_{j}$ stands for the inner product of the color matrices of quarks $i$ and $j$. For the baryon, $\langle\textbf{F}_{i}\cdot\textbf{F}_{j}\rangle=-2/3$. All of the parameters in these formulas are completely consistent with those of our previous works~\cite{F502,F503}.

\subsection{Structure of the wave function }\label{sec2.2}
As is mentioned in Sec.~\ref{sec1}, the $\Sigma_{Q}$, $\Xi_{Q}^{'}$ and $\Omega_{Q}$ baryons belong to the sextet $(\mathbf{6}_{F})$ of the flavor symmetric states.  Their flavor wave functions satisfy the permutation symmetry between the two light quarks~\cite{F403},
\begin{eqnarray}
\begin{aligned}
&\Sigma_{Q}=(uu)Q,~\frac{1}{\sqrt{2}}(ud+du)Q,~(dd)Q, \\
&\Xi_{Q}^{'}=\frac{1}{\sqrt{2}}(us+su)Q,~\frac{1}{\sqrt{2}}(ds+sd)Q,\\
&\Omega_{Q}=(ss)Q.
\end{aligned}
\end{eqnarray}
Here $u$, $d$ and $s$ denote up, down and strange quark, respectively.

\begin{figure}[htbp]
\centering
\includegraphics[width=4.0cm]{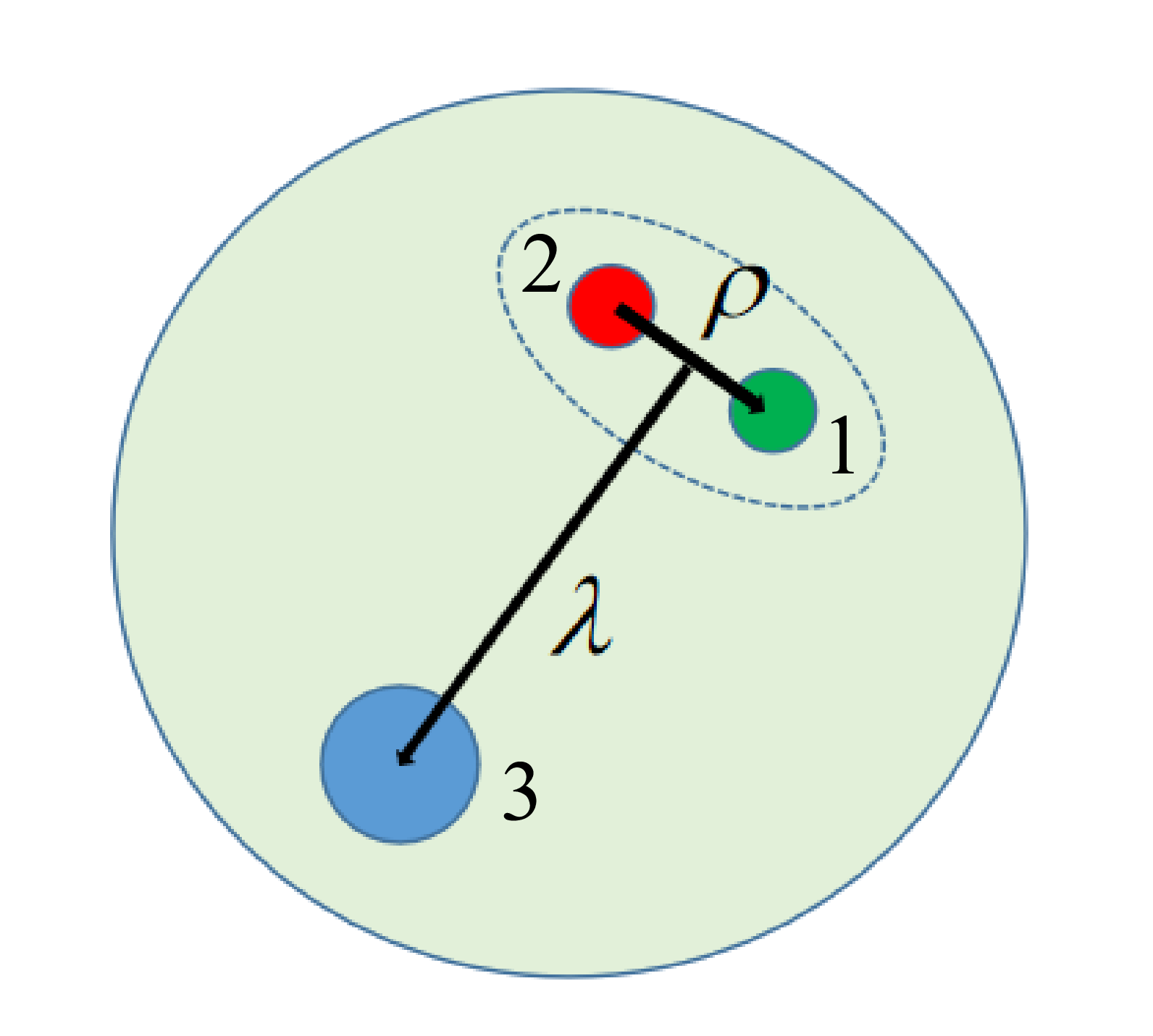}
\caption{The specific Jacobi coordinates (JC-3) for a singly heavy baryon, defining the 3rd quark as the heavy quark. The heavy quark is represented by a bigger ball and the light quarks by two smaller balls. They are numbered for ease of use in calculations.}
\label{f1}
\end{figure}

For describing the internal orbital motion of the singly heavy baryon, we select the specific Jacobi coordinates (named JC-3 for short) as shown in Fig.~\ref{f1}, which is consistent with the above reservation about the flavor wave function naturally.
In this work, the Jacobi coordinates are defined as
\begin{eqnarray}
&\boldsymbol\rho_{i}=\textbf{r}_{jk}=\textbf{r}_{j}-\textbf{r}_{k}, \\
&\boldsymbol\lambda_{i}=\textbf{r}_{i}-\frac{m_{j}\textbf{r}_{j}+m_{k}\textbf{r}_{k}}{m_{j}+m_{k}},
\end{eqnarray}
where $i$, $j$, $k$ = 1, 2, 3(or replace their positions in turn). $\textbf{r}_{i}$ and $m_{i}$ denote the position vector and the mass of the $i$th quark, respectively. So, JC-3 in Fig.~\ref{f1} has the following relationships, $\boldsymbol\rho_{3}\equiv \boldsymbol\rho$ and $\boldsymbol\lambda_{3}\equiv \boldsymbol\lambda$.

Based on the above discussion and the heavy quark effective theory (HQET)~\cite{F303,F304,F403}, the spin and orbital wave function of a baryon state is set to
\begin{eqnarray}
\notag
|l_{\rho}l_{\lambda}Ls_{12}jJM_{J}\rangle &= \{[(|l_{\rho}\ m_{\rho} \rangle |l_{\lambda}\ m_{\lambda} \rangle)_{L}\times(|s_{1}\ m_{s_{1}} \rangle|s_{2}\ m_{s_{2}} \rangle)_{s_{12}}]_{j}\\
& \times|s_{3}\ m_{s_{3}} \rangle \}_{J M_{J}}.
\end{eqnarray}
$l_{\rho}$($l_{\lambda}$), $L$ and $s_{12}$ are the quantum numbers of the relative orbital angular momentum $\textbf{\emph{l}}_{\rho}$ ($\textbf{\emph{l}}_{\lambda}$), total orbital angular momentum $\textbf{\emph{L}}$, and total spin of the light-quark pair $\mathbf{s}_{12}$, respectively. $j$ denotes the quantum number of the coupled angular momentum of $\textbf{\emph{L}}$ and $\textbf{s}_{12}$, so that the total angular momentum $J=j\pm\frac{1}{2}$.
And $M_{J}$ is the 3rd component of $\textbf{\emph{J}}$. Then, the baryon state is simply labeled with $(l_{\rho},l_{\lambda})nL(J^{P})_{j}$, in which $n$ is the quantum number of the radial excitation. It shows that such labeling of quantum states is acceptable, especially $L$ being approximated as a good quantum number~\cite{F501}.
For the $\Sigma_{Q}$, $\Xi_{Q}^{'}$ and $\Omega_{Q}$ baryons, $(-1)^{l_{\rho}+s_{12}}=-1$ should be also guaranteed due to the total antisymmetry of the two light quarks.

\subsection{Improved calculation of Hamiltonian eigenvalues}\label{sec2.3}

The orbital wave function,
\begin{eqnarray}
\notag
(|l_{\rho}\ m_{\rho} \rangle |l_{\lambda}\ m_{\lambda} \rangle)_{L,M_{L}} &&=\sum_{m_{\rho}=-l_{\rho}}^{l_{\rho}}\sum_{m_{\lambda}=-l_{\lambda}}^{l_{\lambda}}
\langle l_{\rho}\ m_{\rho} l_{\lambda}\ m_{\lambda}| l_{\rho}\ l_{\lambda}\ L\ M_{L}\rangle\\
&&\times|l_{\rho}\ m_{\rho} \rangle\otimes |l_{\lambda}\ m_{\lambda} \rangle,
\end{eqnarray}
is expanded in a set of Gaussian basis functions (the Gaussian expansion method, GEM), which forms an approximately complete set in a finite coordinate space and ensures the requirement of high precision calculation~\cite{F601,F6021,F602}.
\begin{eqnarray}
\begin{aligned}
 |l_{\rho} m_{\rho} \rangle\otimes |l_{\lambda} m_{\lambda} \rangle=\sum_{n_{\rho}=1}^{n_{max}}\sum_{n_{\lambda}=1}^{n_{max}}c_{n_{\rho}n_{\lambda}}|n_{\rho} l_{\rho} m_{\rho} \rangle^{G} \otimes |n_{\lambda} l_{\lambda} m_{\lambda} \rangle^{G},
\end{aligned}
\end{eqnarray}
where the Gaussian basis function $|nlm \rangle^{G}$ is commonly written in  position space as
 \begin{eqnarray}
\begin{aligned}
 &\phi^{G}_{nlm}(\textbf{r})=\phi^{G}_{nl}(r)Y_{lm}(\hat{\textbf{r}}),\\
 &\phi^{G}_{nl}(r)=N_{nl}r^{l}e^{-\nu_{n}r^{2}},\\
 &N_{nl}=\sqrt{\frac{2^{l+2}(2\nu_{n})^{l+3/2}}{\sqrt{\pi}(2l+1)!!}},
\end{aligned}
\end{eqnarray}
with
\begin{eqnarray}
\begin{aligned}
& \nu_{n}=\frac{1}{r^{2}_{n}},\\
& r_{n}=r_{1}a^{n-1}\ \ \ (n=1,\ 2,\ ...,\ n_{max}).
\end{aligned}
\end{eqnarray}
$\{r_{1}, a, n_{max}\}$ (or equivalently $\{n_{max},r_{1},r_{n_{max}}\}$) are the Gaussian size
parameters and commonly related to the scale in question~\cite{F602}.  The optimized values of $\{n_{max}=10$, $r_{1}=0.18$ GeV$^{-1}$, $r_{n_{max}}=15$ GeV$^{-1}\}$ are finally selected for the heavy baryons in our works. Details can be found in Refs.~\cite{F502,F503}.

In the calculation of Hamiltonian matrix elements of three-body systems, particularly, when complicated interactions are employed, integrations over all of the radial and angular coordinates become laborious even with the Gaussian basis functions. This process can be simplified by introducing the infinitesimally-shifted Gaussian (ISG) basis functions by
\begin{eqnarray}
\label{e26}
\phi_{nlm}^{G}=N_{nl}r^{l}e^{-\nu_{n}r^{2}}Y_{lm}(\mathbf{\hat{r}})=N_{nl}\lim_{\varepsilon\rightarrow~0}\sum _{k=1}^{k_{max}}C_{lm,k}e^{-\nu_{n}(\mathbf{r}-\varepsilon \mathbf{D}_{lm,k})^{2}},
\end{eqnarray}
where, $Y_{lm}(\mathbf{\hat{r}})$ is replaced by a set of coefficients $C_{lm,k}$ and vectors $\mathbf{D}_{lm,k}$. This is the so-called ISG method, which is used to calculate the Hamiltonian matrix of high orbital excited states so as to effectively avoid the difficulty of the spatial angle integration~\cite{F502,F602}. Especially, the matrix elements, such as  $\langle\alpha|\tilde{G}_{ij}(\mathbf{r}_{ij})|\beta\rangle$, can be conveniently calculated by using the ISG method and the Jacobi coordinate transformation as described below.

For performing the final calculations, the Hamiltonian needs to be expressed as a function of the variables in JC-3, i.e., $H=H(\boldsymbol\rho,\boldsymbol\lambda,\textbf{\emph{l}}_{\rho},\textbf{\emph{l}}_{\lambda})$. With the help of the Jacobi coordinate transformation, the relative coordinates between any two quarks can be represented,
\begin{eqnarray}
\begin{aligned}
& \textbf{r}_{ij}=\alpha_{ij}\boldsymbol\rho+\beta_{ij}\boldsymbol\lambda.
\end{aligned}
\end{eqnarray}
Here, $\alpha_{ij}$ and $\beta_{ij}$ denote the Jacobi coordinate transformation coefficients. So, $\tilde{G}_{ij}(\textbf{r}_{ij})$ and $\tilde{S}_{ij}(\textbf{r}_{ij})$ in Eqs.~(\ref{e8}) and (\ref{e9}) can be calculated in JC-3. Similarly, the momentum-dependent correction factors in Eq.~(\ref{e7}) can be calculated under the corresponding momentum space. Then, $H_{0}$, $\tilde{H}^{conf}_{ij}$ and $\tilde{H}^{hyp}_{ij}$ in Eq.~(\ref{e1}) can be obtained in JC-3.

For the last term in Eq.~(\ref{e1}), $\tilde{H}^{so}$ may take the following approximate form,
\begin{eqnarray}
\begin{aligned}
& \tilde{H}^{so}=\tilde{H}^{so(v)}(\textbf{\emph{l}}_{\rho})+\tilde{H}^{so(v)}(\textbf{\emph{l}}_{\lambda})+\tilde{H}^{so(s)}(\textbf{\emph{l}}_{\rho})+\tilde{H}^{so(s)}(\textbf{\emph{l}}_{\lambda}),
\end{aligned}
\end{eqnarray}
with
\begin{eqnarray}
\label{e20}
 &\tilde{H}^{so(v)}(\textbf{\emph{l}}_{\rho})=\frac{\textbf{s}_{1}\cdot\textbf{\emph{l}}_{\rho}}{2m^{2}_{1}\rho}\frac{\partial\tilde{G}^{so(v)}_{11}}{\partial{\rho}}+
\frac{\textbf{s}_{2}\cdot\textbf{\emph{l}}_{\rho}}{2m^{2}_{2}\rho}\frac{\partial\tilde{G}^{so(v)}_{22}}{\partial{\rho}}+
\frac{(\textbf{s}_{1}+\textbf{s}_{2})\cdot\textbf{\emph{l}}_{\rho}}{m_{1}m_{2}\rho}\frac{\partial\tilde{G}^{so(v)}_{12}}{\partial{\rho}}, \\
\label{e21}
&\tilde{H}^{so(s)}(\textbf{\emph{l}}_{\rho})=-\frac{\textbf{s}_{1}\cdot\textbf{\emph{l}}_{\rho}}{2m^{2}_{1}\rho}\frac{\partial\tilde{S}^{so(s)}_{11}}{\partial{\rho}}-
\frac{\textbf{s}_{2}\cdot\textbf{\emph{l}}_{\rho}}{2m^{2}_{2}\rho}\frac{\partial\tilde{S}^{so(s)}_{22}}{\partial{\rho}},\\
\label{e24}
&\tilde{H}^{so(v)}(\textbf{\emph{l}}_{\lambda})=\frac{\textbf{s}_{3}\cdot\textbf{\emph{l}}_{\lambda}}{2m^{2}_{3}\lambda}\frac{\partial\tilde{G}^{so(v)}_{33}}{\partial{\lambda}}+
\frac{\textbf{s}_{cl.}\cdot\textbf{\emph{l}}_{\lambda}}{2m^{2}_{cl.}\lambda}\frac{\partial\tilde{G}^{so(v)}_{cl.-cl.}}{\partial{\lambda}}+
\frac{(\textbf{s}_{3}+\textbf{s}_{cl.})\cdot\textbf{\emph{l}}_{\lambda}}{m_{3}m_{cl.}\lambda}\frac{\partial\tilde{G}^{so(v)}_{3-cl.}}{\partial{\lambda}}, \\
\label{e25}
&\tilde{H}^{so(s)}(\textbf{\emph{l}}_{\lambda})=-\frac{\textbf{s}_{3}\cdot\textbf{\emph{l}}_{\lambda}}{2m^{2}_{3}\lambda}\frac{\partial\tilde{S}^{so(s)}_{33}}{\partial{\lambda}}-
\frac{\textbf{s}_{cl.}\cdot\textbf{\emph{l}}_{\lambda}}{2m^{2}_{cl.}\lambda}\frac{\partial\tilde{S}^{so(s)}_{cl.-cl.}}{\partial{\lambda}}.
\end{eqnarray}
Here, $cl.$ is used to mark the light-quark cluster (the light-quark pair). $m_{cl.}\approx~m_{1}+m_{2}$ and $\textbf{s}_{cl.}=\textbf{s}_{1}+\textbf{s}_{2}\equiv \mathbf{s}_{12}$ are the approximate mass and the total spin of the cluster, respectively.

Eqs.~(\ref{e20}) and (\ref{e21}) are obtained directly from Eqs.~(\ref{e5}) and (\ref{e6}), by ignoring the mixing of different quantum states.   Eqs.(\ref{e24}) and (\ref{e25}) came from Ref.~\cite{Fp002}, where the terms were named as the ``quasi-two-body spin-orbit forces", and have been used in other theoretical studies~\cite{F405,F407,F603,F604}. It is worth noting that the contributions of the cluster to the spin-orbit interactions (the terms containing $\textbf{s}_{cl.}$) are considered here, while they were neglected in the previous works in the heavy quark limit~\cite{F405,Fp002}. Recently, they were taken into account in Ref.~\cite{F407}. This improvement is important for modifying the spin-orbit energy level splitting, which will be discussed in the next section.

Now, all of the calculations can be performed in JC-3 of Fig.\ref{f1}. And we can obtain the Hamiltonian eigenvalue of each baryon state $(l_{\rho},l_{\lambda})nL(J^{P})_{j}$ mentioned in Sec.~\ref{sec2} B. More calculation details have been described in our previous works~\cite{F502,F503}.

\section{Results and discussions}\label{sec3}

\subsection{Fine structures and Retest of the HQD mechanism }\label{sec3.1}

For the $L$-wave excitation with $\textbf{\emph{L}}$=$\textbf{\emph{l}}_{\rho}$+$\textbf{\emph{l}}_{\lambda}$, there are an infinite number of orbital excitation modes.
Taking $L=1$ as an example, the excitation modes $(l_{\rho},l_{\lambda})_{L}$ are $(1,0)_{1}$, $(0,1)_{1}$, $(1,1)_{1}$, $(2,1)_{1}$, $(1,2)_{1}$, $(2,2)_{1}$, and so on. We assume that the excitation mode with the lowest energy level is the most stable and has the greatest probability of being observed experimentally, which dominates the structure of the excitation spectrum. This assumption is summarized as the HQD approximation~\cite{F501}.

\begin{figure}[htbp]
\centering
\includegraphics[width=8.5cm]{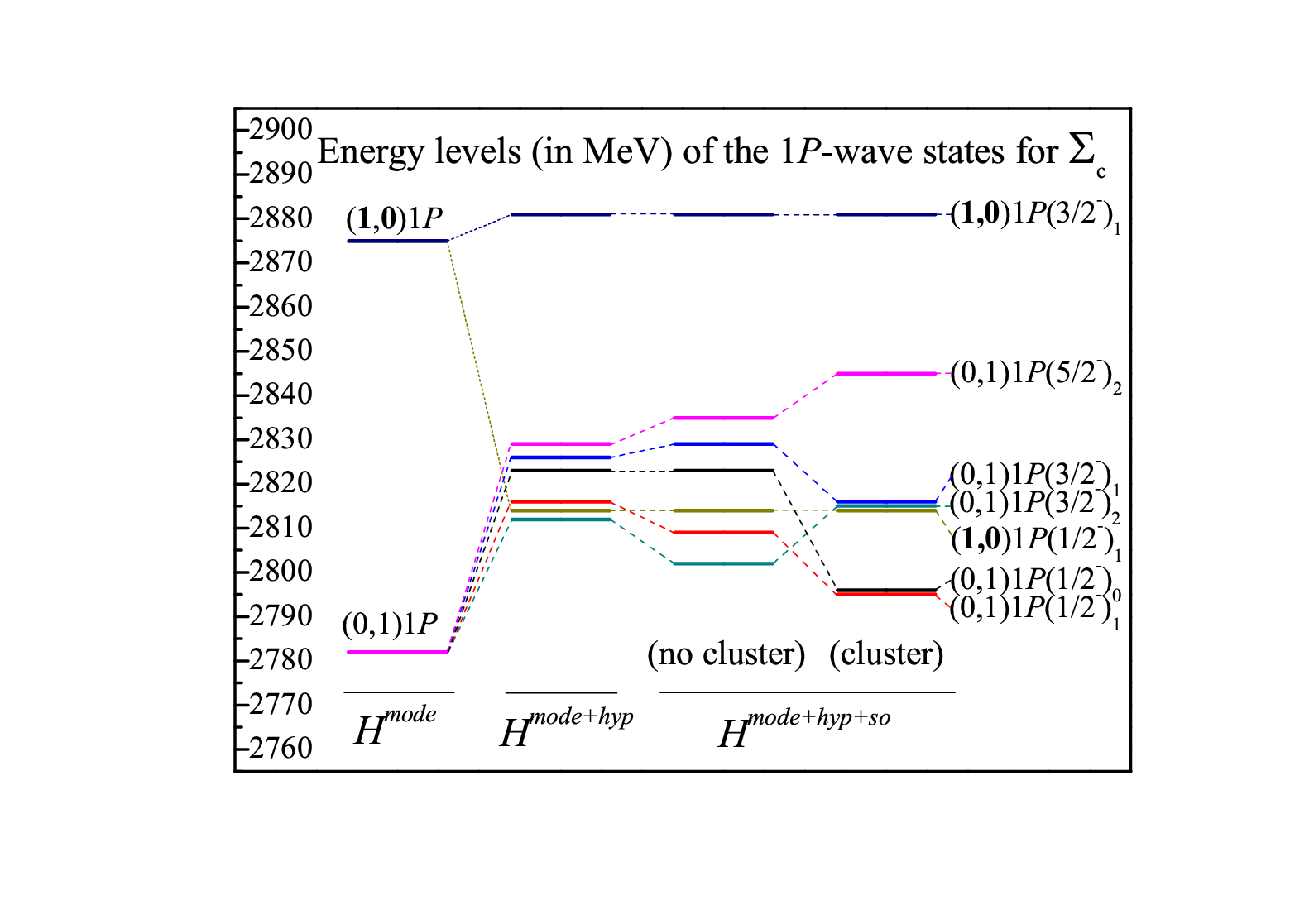}
\caption{The contributions of the Hamiltonian terms to the energy levels (in MeV) are presented. $H^{mode}=H_{0}+\tilde{H}^{conf}$ determines the energy levels of the orbital excitation modes~\cite{F501}. The hyperfine and spin-orbit interactions enhance the energy level splitting to different extents for $(0,1)_{1}$ and $(1,0)_{1}$ excitation modes, respectively. For the $(0,1)_{1}$ mode, the contribution of the light-quark cluster modifies the energy levels significantly. This leads to the emergency of the reasonable fine structure, where the states with same $J^{P}$ values are closed together and the states with different $J^{P}$ values separate from each other. For the $(1,0)_{1}$ mode, its energy level splitting is not caused by $\tilde{H}^{so}$ owing to $s_{12}=0$ and $l_{\lambda}=0$ there. }
\label{f2}
\end{figure}

In this work, we consider the contribution of the light-quark cluster to the spin-orbit interaction and improve the corresponding calculations. Taking the $1P$-wave states of the $\Sigma_{c}$ baryon as an example, this improvement modifies the energy level splitting largely, as shown in Fig.\ref{f2}. Specially, this leads to the emergence of the ``reasonable'' fine structures.

\begin{figure*}[htbp]
\centering
\includegraphics[width=16.5cm]{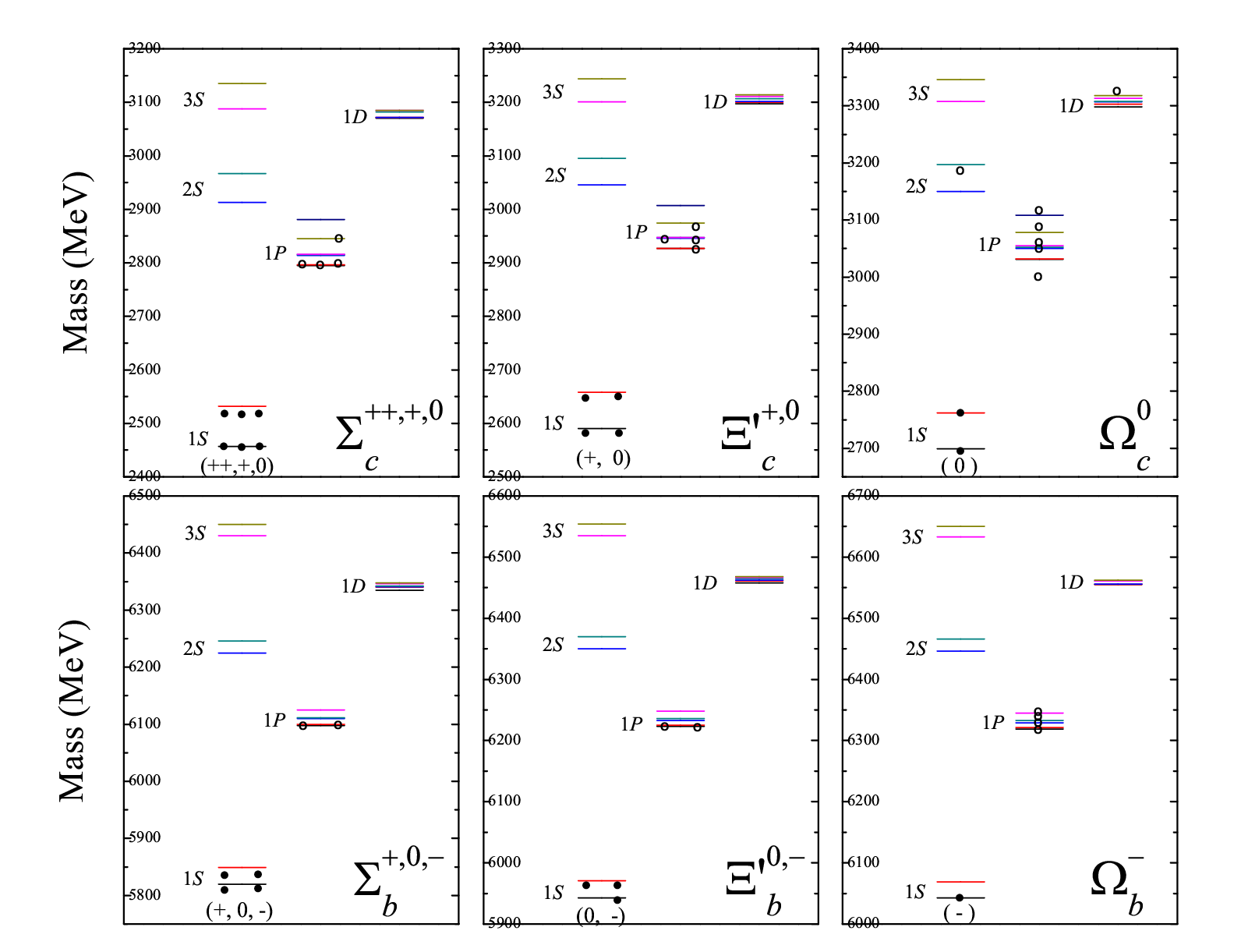}
\caption{Calculated spectra of singly heavy baryons and the relevant experimental data~\cite{F201,F209,F203,F204,F206}. `++',`+', `0' and `-' in the brackets indicate the charged states of baryons. The solid black circles denote the baryons with confirmed spin-parity values, and the open circles are the ones whose spin-parities have not been identified. }
\label{f3}
\end{figure*}

Based on the above theoretical methods and the HQD approximation, we obtain the mass values and the corresponding root mean square radii for the $1S$-, $2S$-, $3S$-, $1P$- and $1D$-wave states, and list the results of the $\Sigma_{c(b)}$, $\Xi_{c(b)}^{'}$ and $\Omega_{c(b)}$ families in Tables~\ref{ta2}, ~\ref{ta3} and~\ref{ta4}, respectively. In order to be more intuitive, their energy level structures are shown in Fig.\ref{f3}.
For the bottom baryons, there are a total of five $1P$-wave states. They all belong to the $\lambda$-mode ($l_{\rho}=0,l_{\lambda}\neq0$) and are divided into three groups in order of their energy levels and spin-parity values. From the bottom half of Fig.\ref{f3}, one can see their clear fine structures of the $1P$-wave states.

The case of the charm baryons is a little more complicated. As discussed in Refs.~\cite{F327,F501}, $c$ quark might be not so heavy and hardly treated as a heavy quark. The energy level structure of the five $1P$-wave states in the $\lambda$-mode is the same as that of the bottom baryons. While, $(1,0)1P(\frac{1}{2})_{1}$ state, as one of the $\rho$-mode ($l_{\rho}\neq0,l_{\lambda}=0$) states, intrudes the energy region of the $\lambda$-mode and hovers near the $(0,1)1P(\frac{3}{2})_{1,2}$ states, as shown in Tables~\ref{ta2}, ~\ref{ta3} and~\ref{ta4} (or Fig.\ref{f2}). The other $\rho$-mode state $(1,0)1P(\frac{3}{2})_{1}$ occupies an energy level by itself. Thus, there are clearly four energy levels separated from each other for the $1P$-wave states in these charm baryons, as shown in Fig.\ref{f3}. So, the two excitation modes coexist in the $1P$-wave states and present an interesting fine structure.

The HQD mechanism in the orbital excitation was proposed and investigated in Ref.~\cite{F501}.
This mechanism means that the excitation mode with lower energy levels is always associated with the heavy quark(s), and the splitting of the energy levels is suppressed by the heavy quark(s) as well. In other words, the heavy quarks dominate the orbital excitation of singly and doubly heavy baryons, and determine the structures of their excitation spectra.
As shown in Tables~\ref{ta2}, ~\ref{ta3} and~\ref{ta4}, the lower excitation modes come mainly from the $\lambda$-modes, which is associated with the heavy quark. But there are exceptions for the $1P$-wave orbital excitation of $\Sigma_{c}$, $\Xi_{c}^{'}$ and $\Omega_{c}$ baryons, where the $\rho$-modes intrude the energy region of the $\lambda$-modes. In addition, one can see the energy level splitting of the $\lambda$-mode is certainly suppressed by the heavy quark. Taking the $(0,1)1P(\frac{1}{2},\frac{3}{2})_{1}$ doublet states as an example, their energy difference is 21 MeV for the $\Sigma_{c}$ baryons, but 10 MeV for the $\Sigma_{b}$ baryons, as shown in Table~\ref{ta2}.
So, we conclude that the HQD mechanism is generally effective. But for the $1P$-wave orbital excitation of the $\Sigma_{c}$, $\Xi_{c}^{'}$ and $\Omega_{c}$ baryons, it is slightly broken, since $c$ quark is not heavy enough.

In 1978, Isgur and Karl made a discussion about the symmetry breaking in baryons, where a similar conclusion was obtained for the strange baryons~\cite{Fp004}. The HQD mechanism and its breaking could be viewed as an extension of that issue to the heavy baryons.

\subsection{Assignments of the observed baryons}\label{sec3.2}

Since the appearance of the fine structures in the calculated excitation spectra, it becomes possible to theoretically distinguish between the excited heavy baryons appearing in groups.

\emph{(1) $\Sigma_{c}$ and $\Sigma_{b}$ baryons.}
 The spin-parity of the $\Sigma_{c}(2800)^{++,+,0}$ baryons is still unconfirmed in experiment. In Table~\ref{ta2}, one can see their mass values fall exactly on the calculated value of the $(0,1)1P(\frac{1}{2}^{-})_{1}$ or $(0,1)1P(\frac{1}{2}^{-})_{1}$ state. So, these two predicted states might be the ideal candidates. Since they have the same spin-parity in theory, we can preliminarily judge that the spin-parity $J^{P}$ of the $\Sigma_{c}(2800)^{++,+,0}$ baryons should be $\frac{1}{2}^{-}$.

 The related theoretical works about mass$(J^{P})$ (Mass values are in MeV) of $\Sigma_{c}(2800)^{++,+,0}$ are collected as follows: $2770(\frac{1}{2}^{-})$ or $2805(\frac{3}{2}^{-})$~\cite{F402}, $2713(\frac{1}{2}^{-})$ or $2773(\frac{3}{2}^{-})$~\cite{F405}, $2768(\frac{1}{2}^{-})$ or $2763(\frac{3}{2}^{-})$~\cite{F403}, $2791(\frac{1}{2}^{-})$ or $2791(\frac{3}{2}^{-})$~\cite{Fp005}, $2798(\frac{3}{2}^{-})$~\cite{F407}, $2809(\frac{1}{2}^{-})$ or $2802(\frac{3}{2}^{-})$~\cite{F502}, and $2820(\frac{3}{2}^{-})$~\cite{Fp003}. It can be seen that most predictions give two possible assignments. It is difficult to accurately identify the locations of these baryons in the fine structure energy range.

 $\Sigma_{c}(2846)^{0}$ was observed by BaBar collaboration, with $m=2846\pm8\pm10$ MeV, which is inconsistent with the measurement of $\Sigma_{c}(2800)^{0}$~\cite{F209}. Table~\ref{ta2} shows its mass value is very close to that of the $(0,1)1P(\frac{5}{2}^{-})_{2}$ state. So, it can be preliminarily assigned to this state~\cite{F411}. Then, its spin is up to $\frac{5}{2}$, which means it might be extremely unstable and its decay width would be large. This feature agrees with the measured width $86^{+33}_{-22}\pm12$ MeV~\cite{F209}.

 Some theoretical works assigned $\Sigma_{c}(2846)^{0}$ to the radial excited state $2S(\frac{1}{2}^{+})$~\cite{F402,F405}, but their predicted mass values are too large. While the predicted mass$[nL (J^{P})]$ is 2850$[2S(\frac{1}{2}^{+})]$ in Ref.~\cite{F407}. In Ref.~\cite{Fp003}, $\Sigma_{c}(2846)^{0}$ was assigned to the $(\frac{1}{2}^{-})$ or $(\frac{5}{2}^{-})$ state.

 The case of $\Sigma_{b}(6097)^{+,-}$ is similar to that of $\Sigma_{c}(2800)^{++,+,0}$. So, we can safely conclude that the $J^{P}$ of $\Sigma_{b}(6097)^{+,-}$ is likely to be $\frac{1}{2}^{-}$. And they should be assigned to the $(0,1)1P(\frac{1}{2}^{-})_{0}$ or $(0,1)1P(\frac{1}{2}^{-})_{1}$ state.
 In some other studies, $\Sigma_{b}(6097)^{+,-}$ is considered to be the $(\frac{1}{2}^{-})$ or $(\frac{3}{2}^{-})$ state~\cite{F403,F405,F502,Fp003}. While, it is assigned to the $(\frac{3}{2}^{-})$ or $(\frac{5}{2}^{-})$ state in Ref.~\cite{F324,Fp006}.

\emph{(2) $\Xi_{c}^{'}$ and $\Xi_{b}^{'}$ baryons.}
 A charged $\Xi_{c}(2930)^{+}$ baryon was observed by Belle collaboration in 2018~\cite{F204}. Later, $\Xi_{c}(2923)^{0}$, $\Xi_{c}(2939)^{0}$ and $\Xi_{c}(2964)^{0}$ states were observed with a large significance by LHCb collaboration~\cite{F203}. As shown in Table~\ref{ta3}, we preliminarily assign the $\Xi_{c}(2923)^{0}$, $\Xi_{c}(2939)^{0}$ and $\Xi_{c}(2964)^{0}$ baryons to the $(0,1)1P(\frac{1}{2}^{-})_{0}$ [or $(0,1)1P(\frac{1}{2}^{-})_{1}$],  $(0,1)1P(\frac{3}{2}^{-})_{1}$ [or $(0,1)1P(\frac{3}{2}^{-})_{2}$] and $(0,1)1P(\frac{5}{2}^{-})_{2}$ states, respectively, for the exact correspondence of their mass values. The decay modes of $\Xi_{c}(2930)^{+}$ and $\Xi_{c}(2939)^{0}$  are quite different~\cite{F203}. They should not belong to the same state. So, we consider the $\Xi_{c}(2930)^{+}$ baryon should be the intrusion state $(1,0)1P(\frac{1}{2}^{-})_{1}$.

As shown in Table~\ref{ta3}, the $\Xi_{b}(6227)^{0,-}$ baryons might be assigned to the $(0,1)1P(\frac{1}{2}^{-})_{0}$ or $(0,1)1P(\frac{1}{2}^{-})_{1}$ state. So, their spin-parity should be $\frac{1}{2}^{-}$. There are great similarities between $\Xi_{b}(6227)^{0,-}$ and $\Sigma_{b}(6097)^{+,-}$ as shown in Fig.\ref{f3}.

In another systematical work about $\Xi_{c}^{'}$ and $\Xi_{b}^{'}$ baryons~\cite{Fp007}, these baryons were assigned in a different way, i.e., $\Xi_{c}(2923)^{0}~\rightarrow~^{4}\lambda_{1/2^{-}}$, $\Xi_{c}(2939)^{0} \rightarrow~ ^{2}\lambda_{3/2^{-}}$, $\Xi_{c}(2964)^{0} \rightarrow~^{4}\lambda_{3/2^{-}}$, $\Xi_{c}(2930)^{+} \rightarrow~^{2}\lambda_{3/2^{-}}$, and $\Xi_{b}(6227)^{0,-} \rightarrow~^{4}\lambda_{5/2^{-}}$.

\emph{(3) $\Omega_{c}$ and $\Omega_{b}$ baryons.}
$\Omega_{c}(3185)^{0}$ might be assigned to the radial excited state $2S(\frac{3}{2}^{+})$,  because they have the closest mass values. As shown in Table~\ref{ta4}, there are five baryons in the $1P$-wave states. According to the mass values, they might be assigned as follows:
$\Omega_{c}(3050)^{0}\rightarrow (0,1)1P(\frac{3}{2}^{-})_{1}$ (or $(0,1)1P(\frac{3}{2}^{-})_{2}$), $\Omega_{c}(3065)^{0}\rightarrow (1,0)1P(\frac{1}{2}^{-})_{1}$ (an  intrusion state), $\Omega_{c}(3090)^{0}\rightarrow (0,1)1P(\frac{5}{2}^{-})_{2}$ and $\Omega_{c}(3120)^{0}\rightarrow (1,0)1P(\frac{3}{2}^{-})_{1}$ (the $\rho$-excitation mode). Here, the successful assignment of $\Omega_{c}(3120)^{0}$ implies that our treatment of the energy level structure is correct~\cite{F2021,F411}. We temporarily assign $\Omega_{c}(3000)^{0}$ to the $(0,1)1P(\frac{1}{2}^{-})_{0}$ state. This assignment is consistent with that in Ref.~\cite{F411}. Unfortunately, $\Omega_{c}(3000)^{0}$ can not really be assigned very well in this spectrum, because its mass is so different from the predicted ones. Or, it might be not a pure baryon state~\cite{F322,F3221,F323,F3231}. $\Omega_{c}(3327)^{0}$ should belong to the $1D$-wave states. Cautiously, it is hard to assign this baryon exactly, due to the energy level quasi-degeneracy of these $1D$-wave states.

These four $1P$-wave $\Omega_{b}$ baryons can be well assigned in order of their mass values, as shown in Table~\ref{ta4}. Their assignments are as follows: $\Omega_{b}(6316)^{-}\rightarrow (0,1)1P(\frac{1}{2}^{-})_{0}$ (or $(0,1)1P(\frac{1}{2}^{-})_{1}$), $\Omega_{b}(6330)^{-}\rightarrow (0,1)1P(\frac{3}{2}^{-})_{1}$, $\Omega_{b}(6340)^{-}\rightarrow (0,1)1P(\frac{3}{2}^{-})_{2}$ and $\Omega_{b}(6350)^{-}\rightarrow (0,1)1P(\frac{5}{2}^{-})_{2}$.
By comparison, these excited $\Omega_{Q}$ baryons were systematically analyzed in Ref.~\cite{Fp003}, which gives almost the same result as ours.

\subsection{Reliability analysis of the calculation }\label{sec3.3}

As is mentioned in Sec.~\ref{sec1}, the $\Lambda_{Q}$ and $\Xi_{Q}$ baryons belong to the $\mathbf{\bar{3}}_{F}$ sector, and $(-1)^{l_{\rho}+s_{12}}=1$ should be guaranteed, which is different from the situation of the $\mathbf{6}_{F}$ sector (see Sec.~\ref{sec2} B). Their orbital excitation is dominated only by the $\lambda$-mode, i.e., $l_{\rho}=0$ and $s_{12}=0$ (corresponding to $s_{cl.}=0$ in this paper). So, the improved calculation of the spin-orbital terms has no effect on the spectra of the $\Lambda_{Q}$ and $\Xi_{Q}$ families, which have been presented in our previous papers~\cite{F502,F503}.

In those papers, the predicted excited $\Xi_{Q}$ states were accurately verified by later experiments, such as $\Xi_{b}(6095)^{0}$ and $\Xi_{b}(6087)^{0}$~\cite{F207}. Very recently, LHCb collaboration determined the spin-parity of the $\Xi_{c}(3055)^{+,0}$ baryons for the first time~\cite{F214}, which are exactly the same as our predictions~\cite{F503,F501}. The successful description of the experimental data reflects the systematic and reliable nature of the method.

The $\Sigma_{c(b)}$, $\Xi '_{c(b)}$ and $\Omega_{c(b)}$ families belong to the $\mathbf{6}_{F}$ sector. Their spectral structure are similar to each other. We can judge the reliability of the calculation by comparing these excited baryons  systematically. If the $\Xi_{b}(6227)^{0,-}$ was regarded as the strange partners of the $\Sigma_{b}(6097)^{+,-}$, the experimental observations of the $\Xi_{b}(6227)^{0,-}$ should be in the same quantum states as the $\Sigma_{b}(6097)^{+,-}$. So, it is credible that $\Xi_{b}(6227)^{0,-}$ and $\Sigma_{b}(6097)^{+,-}$ are both in the lowest excited $1P$-wave states with $J^{P}=\frac{1}{2}^{-}$, according to our calculation.

In Ref.~\cite{F4071}, the following chain was found by analyzing the universal behavior of their mass gaps,
 \begin{eqnarray}
\begin{aligned}
\Sigma_{c}(2846)^{0}\leftrightarrow\Xi_{c}^{'}(2964)^{0}\leftrightarrow\Omega_{c}(3090)^{0} ,
\end{aligned}
\end{eqnarray}
which implies that these baryons are in the same quantum state. As is shown in Fig.~\ref{f3}, these three baryons are indeed lying on the green lines. They are all in the $(0,1)1P(\frac{5}{2}^{-})_{2}$ states as shown in Tables~\ref{ta2}, ~\ref{ta3} and~\ref{ta4}. This is mutually confirmed with the finding of Ref.~\cite{F4071}.

In addition, the $\Xi_{c}(2930)^{0}$ was observed by Belle collaboration in 2017~\cite{F208}. While the observation from LHCb collaboration indicates that the $\Xi_{c}(2930)^{0}$ might be an overlap of two narrower states, such as the $\Xi_{c}(2923)^{0}$ and $\Xi_{c}(2939)^{0}$ baryons~\cite{F203}. So, there are only the four excited $\Xi_{c}^{'}$ baryons listed in Table~\ref{ta3}. They should have a structural correspondence with the excited $\Omega_{c}$ baryons listed in
Table~\ref{ta4}. The absence of the excited $\Xi_{c}^{'}$ baryon in the $(1,0)1P(\frac{3}{2}^{-})_{1}$ state may be attributed to its large decay width, which needs a further theoretical study.

For the excited $\Omega_{c}$ baryons, the predicted fine structure of the $1P$-wave states corresponds well with the experimental observations (see Fig.~\ref{f3}). From this perspective, the $\Omega_{c}(3000)^{0}$ is most likely to be interpreted as the $(0,1)1P(\frac{1}{2}^{-})_{0}$ state, even though their mass difference is relatively large.

Since the spectral structures of these baryon families are similar, the systematical analysis can enhance the reliability of the theoretical predictions.
 From Fig.\ref{f3}, it can be seen that the theoretical results match well with the experimental data. This implies that the method used in this work is reliable and the theoretical prediction is credible. In fact, this method can get exactly the same results as those in our previous works on the $\Lambda_{c(b)}$ and $\Xi_{c(b)}$ baryons.  So, the unified description and precise calculation of all singly heavy baryon spectra have been achieved.

 On the other hand, checking the number of the observed excited states from left to right in Fig.\ref{f3}, one can find that the number of the $1P$-wave excited baryons increases gradually, sometimes suddenly. This suggests that there is an evolutionary law in the structure of the singly heavy baryon spectra.

\begin{table*}[htbp]
\begin{ruledtabular}\caption{Calculated $\langle r_{\rho}^{2}\rangle^{1/2}$, $\langle r_{\lambda}^{2}\rangle^{1/2}$ (in fm) and mass values (in MeV) for the $1S$-, $2S$-, $3S$-, $1P$- and $1D$-wave states of the $\Sigma_{c}$ and $\Sigma_{b}$ baryons. The orbital excitation states of the $\rho$-mode are marked in bold type. The experimental data are also listed for comparison, taken by their isospin averages.}
\label{ta2}
\begin{tabular}{c c c c c c c c c c c}
\multirow{2}{*}{$(l_{\rho},l_{\lambda})nL(J^{P})_{j}$}& \multicolumn{4}{c}{$\Sigma_{c}$}   &\multicolumn{4}{c}{$\Sigma_{b}$} \\\cline{2-5} \cline{6-9}
 & $\langle r_{\rho}^{2}\rangle^{1/2}$ & $\langle r_{\lambda}^{2}\rangle^{1/2}$ & $M_{cal.}$ & Baryon/$M_{exp.}/J^{P}_{exp.}$ & $\langle r_{\rho}^{2}\rangle^{1/2}$ & $\langle r_{\lambda}^{2}\rangle^{1/2}$ & $M_{cal.}$ & Baryon/$M_{exp.}/J^{P}_{exp.}$ \\ \hline
$(0,0)1S(\frac{1}{2}^{+})_{1}$ & 0.617  & 0.446 & 2457 & $\Sigma_{c}(2455)^{++,+,0}$/$\sim$2453/$\frac{1}{2}^{+}$~\cite{F201} & 0.633  & 0.430 & 5820 & $\Sigma_{b}^{+,-}$/$\sim$5813/$\frac{1}{2}^{+}$~\cite{F201} \\
$(0,0)1S(\frac{3}{2}^{+})_{1}$ & 0.642  & 0.494 & 2532 & $\Sigma_{c}(2520)^{++,+,0}$/$\sim$2518/$\frac{3}{2}^{+}$~\cite{F201} & 0.644  & 0.451 & 5849 & $\Sigma_{b}^{*+,-}$/$\sim$5833/$\frac{3}{2}^{+}$~\cite{F201} \\
$(0,0)2S(\frac{1}{2}^{+})_{1}$ & 0.855  & 0.721 & 2913 & - & 0.776  & 0.714 & 6225 & - \\
$(0,0)2S(\frac{3}{2}^{+})_{1}$ & 0.832  & 0.785 & 2967 & - & 0.769  & 0.734 & 6246 & -\\
$(0,0)3S(\frac{1}{2}^{+})_{1}$ & 0.936  & 0.703 & 3088 & - & 1.019  & 0.594 & 6430 & -  \\
$(0,0)3S(\frac{3}{2}^{+})_{1}$ & 0.993  & 0.706 & 3135 & - & 1.041  & 0.600 & 6450 & -  \\\\
$(0,1)1P(\frac{1}{2}^{-})_{1}$ & 0.663  & 0.654 & 2795 &$\Sigma_{c}(2800)^{++,+,0}$/$\sim$2800/$?^{?}$~\cite{F201} & 0.659  & 0.605 & 6100 & $\Sigma_{b}(6097)^{+,-}$/$\sim$6097/$?^{?}$~\cite{F201}\\
$(0,1)1P(\frac{1}{2}^{-})_{0}$ & 0.661  & 0.648 & 2796 & - & 0.655  & 0.595 & 6098 & - \\\\
$\mathbf{(1,0)}1P(\frac{1}{2}^{-})_{1}$ & 0.847  & 0.464 & 2814 & - & - &- & - & - \\
$(0,1)1P(\frac{3}{2}^{-})_{2}$ & 0.676  & 0.689 & 2815 & - & 0.671  & 0.635 & 6111 & - \\
$(0,1)1P(\frac{3}{2}^{-})_{1}$ & 0.671  & 0.674 & 2816 & - & 0.676  & 0.689 & 6110 & - \\\\
$(0,1)1P(\frac{5}{2}^{-})_{2}$ & 0.688  & 0.720 & 2845 &  $\Sigma_{c}(2846)^{0}$/$\sim$2846/$?^{?}$~\cite{F209}  & 0.677  & 0.649 & 6125 & - \\\\
$\mathbf{(1,0)}1P(\frac{3}{2}^{-})_{1}$ & 0.879  & 0.514 & 2881 & - & - & - & - & -  \\\\
$(0,2)1D(\frac{1}{2}^{+})_{1}$ & 0.683  & 0.873 & 3070 & - & 0.670  & 0.766 & 6341 & - \\
$(0,2)1D(\frac{3}{2}^{+})_{2}$ & 0.689  & 0.858 & 3072 & - & 0.676  & 0.784 & 6340 & -  \\
$(0,2)1D(\frac{5}{2}^{+})_{3}$ & 0.697  & 0.887 & 3072 & - & 0.684  & 0.809 & 6335 & - \\
$(0,2)1D(\frac{3}{2}^{+})_{1}$ & 0.687  & 0.852 & 3082 & - & 0.673  & 0.773 & 6348 & - \\
$(0,2)1D(\frac{5}{2}^{+})_{2}$ & 0.693  & 0.873 & 3085 & - & 0.679  & 0.791 & 6346 & - \\
$(0,2)1D(\frac{7}{2}^{+})_{3}$ & 0.712  & 0.905 & 3085 & - & 0.687  & 0.818 & 6342 & -\\
\end{tabular}
\end{ruledtabular}
\end{table*}

\begin{table*}[htbp]
\begin{ruledtabular}\caption{Same as Table~\ref{ta2}, but for the $\Xi'_{c}$ and $\Xi'_{b}$ baryons.  }
\label{ta3}
\begin{tabular}{c c c c c c c c c }
\multirow{2}{*}{$(l_{\rho},l_{\lambda})nL(J^{P})_{j}$}& \multicolumn{4}{c}{$\Xi'_{c}$}   &\multicolumn{4}{c}{$\Xi'_{b}$} \\\cline{2-5} \cline{6-9}
 & $\langle r_{\rho}^{2}\rangle^{1/2}$ & $\langle r_{\lambda}^{2}\rangle^{1/2}$ & $M_{cal.}$ & Baryon/$M_{exp.}/J^{P}_{exp.}$ & $\langle r_{\rho}^{2}\rangle^{1/2}$ & $\langle r_{\lambda}^{2}\rangle^{1/2}$ & $M_{cal.}$ & Baryon/$M_{exp.}/J^{P}_{exp.}$ \\ \hline
$(0,0)1S(\frac{1}{2}^{+})_{1}$ & 0.590  & 0.431 & 2590 & $\Xi'^{+,0}_{c}$ /$\sim$2578/$\frac{1}{2}^{+}$~\cite{F201} & 0.604  & 0.411 & 5943 & $\Xi_{b}(5935)^{-}$/$\sim$5935/$\frac{1}{2}^{+}$~\cite{F201} \\
$(0,0)1S(\frac{3}{2}^{+})_{1}$ & 0.611  & 0.476 & 2658 & $\Xi_{c}(2645)^{+,0}$/$\sim$2645/$\frac{3}{2}^{+}$~\cite{F201}& 0.614  & 0.431 & 5971 & $\Xi_{b}(5955)^{0,-}$/$\sim$5954/$\frac{3}{2}^{+}$~\cite{F201}  \\
$(0,0)2S(\frac{1}{2}^{+})_{1}$ & 0.821  & 0.705 & 3046 & -  & 0.741  & 0.697 & 6350 & -\\
$(0,0)2S(\frac{3}{2}^{+})_{1}$ & 0.801  & 0.763 & 3095 & - & 0.735  & 0.716 & 6370 & - \\
$(0,0)3S(\frac{1}{2}^{+})_{1}$ & 0.918  & 0.671 & 3201 & - & 0.998  & 0.559 & 6535 & - \\
$(0,0)3S(\frac{3}{2}^{+})_{1}$ & 0.968  & 0.676 & 3244 & - & 1.017  & 0.566 & 6554 & - \\\\
$(0,1)1P(\frac{1}{2}^{-})_{0}$ & 0.637  & 0.637 & 2927 & $\Xi_{c}(2923)^{0}$/$\sim$2923/$?^{?}$~\cite{F203} & 0.632  & 0.581 & 6223 & $\Xi_{b}(6227)^{0,-}$/$\sim$6227/$?^{?}$~\cite{F201} \\
$(0,1)1P(\frac{1}{2}^{-})_{1}$ & 0.638  & 0.642 & 2927 & - & 0.635  & 0.589 & 6225 & -\\\\
$(0,1)1P(\frac{3}{2}^{-})_{1}$ & 0.645  & 0.660 & 2946 & $\Xi_{c}(2939)^{0}$/$\sim$2939/$?^{?}$~\cite{F203} & 0.638  & 0.597 & 6233 & - \\
$(0,1)1P(\frac{3}{2}^{-})_{2}$ & 0.648  & 0.670 & 2947 & - & 0.643  & 0.613 & 6236 & - \\
$\mathbf{(1,0)}1P(\frac{1}{2}^{-})_{1}$ & 0.822  & 0.453 & 2947 & $\Xi_{c}(2930)^{+}$/$\sim$2942/$?^{?}$~\cite{F204}  & - & - & - & -  \\\\
$(0,1)1P(\frac{5}{2}^{-})_{2}$ & 0.658  & 0.698 & 2974 & $\Xi_{c}(2964)^{0}$/$\sim$2965/$?^{?}$~\cite{F203} & 0.648  & 0.626 & 6248 & - \\\\
$\mathbf{(1,0)}1P(\frac{3}{2}^{-})_{1}$ & 0.849  & 0.499 & 3007 & - & - & - & - & -\\\\
$(0,2)1D(\frac{1}{2}^{+})_{1}$ & 0.661  & 0.827 & 3197 & -  & 0.650  & 0.752 & 6462 & -\\
$(0,2)1D(\frac{3}{2}^{+})_{2}$ & 0.666  & 0.844 & 3200 & - & 0.654  & 0.767 & 6461 & -\\
$(0,2)1D(\frac{5}{2}^{+})_{3}$ & 0.671  & 0.866 & 3202 & - & 0.660  & 0.786 & 6458 & - \\
$(0,2)1D(\frac{3}{2}^{+})_{1}$ & 0.665  & 0.841 & 3207 & - & 0.652  & 0.759 & 6468 & -\\
$(0,2)1D(\frac{5}{2}^{+})_{2}$ & 0.669  & 0.858 & 3211 & - & 0.774  & 0.689 & 6467 & -\\
$(0,2)1D(\frac{7}{2}^{+})_{3}$ & 0.675  & 0.883 & 3214 & - & 0.662  & 0.794 & 6465 & - \\
\end{tabular}
\end{ruledtabular}
\end{table*}

\begin{table*}[htbp]
\begin{ruledtabular}\caption{Same as Table~\ref{ta2}, but for the $\Omega_{c}$ and $\Omega_{b}$ baryons. }
\label{ta4}
\begin{tabular}{c c c c c c c c c }
\multirow{2}{*}{$(l_{\rho},l_{\lambda})nL(J^{P})_{j}$}& \multicolumn{4}{c}{$\Omega_{c}$}   &\multicolumn{4}{c}{$\Omega_{b}$} \\\cline{2-5} \cline{6-9}
 & $\langle r_{\rho}^{2}\rangle^{1/2}$ & $\langle r_{\lambda}^{2}\rangle^{1/2}$ & $M_{cal.}$ & Baryon/$M_{exp.}/J^{P}_{exp.}$ & $\langle r_{\rho}^{2}\rangle^{1/2}$ & $\langle r_{\lambda}^{2}\rangle^{1/2}$ & $M_{cal.}$ & Baryon/$M_{exp.}/J^{P}_{exp.}$ \\ \hline
$(0,0)1S(\frac{1}{2}^{+})_{1}$ & 0.555  & 0.414 & 2699 & $\Omega_{c}^{0}$/$\sim$2695/$\frac{1}{2}^{+}$~\cite{F201} & 0.565  & 0.392 & 6043 & $\Omega_{b}^{-}$/$\sim$6045/$\frac{1}{2}^{+}$~\cite{F201}  \\
$(0,0)1S(\frac{3}{2}^{+})_{1}$ & 0.575  & 0.455 & 2762 & $\Omega_{c}(2770)^{0}$/$\sim$2766/$\frac{3}{2}^{+}$~\cite{F201}  & 0.575  & 0.410 & 6069 & - \\
$(0,0)2S(\frac{1}{2}^{+})_{1}$ & 0.782  & 0.681 & 3150 & - & 0.706  & 0.670 & 6446 & -\\
$(0,0)2S(\frac{3}{2}^{+})_{1}$ & 0.770  & 0.730 & 3197 & $\Omega_{c}(3185)^{0}$/$\sim$3185/$?^{?}$~\cite{F206} & 0.702  & 0.688 & 6466 & - \\
$(0,0)3S(\frac{1}{2}^{+})_{1}$ & 0.880  & 0.655 & 3308 & - & 0.954  & 0.548 & 6633 & -\\
$(0,0)3S(\frac{3}{2}^{+})_{1}$ & 0.923  & 0.664 & 3346 & - & 0.972  & 0.555 & 6650 & -\\\\
$(0,1)1P(\frac{1}{2}^{-})_{0}$ & 0.605  & 0.616 & 3031 &$\Omega_{c}(3000)^{0}$/$\sim$3000/$?^{?}$~\cite{F201}  & 0.598  & 0.558 & 6319 & $\Omega_{b}(6316)^{-}$/$\sim$6315/$?^{?}$~\cite{F201} \\
$(0,1)1P(\frac{1}{2}^{-})_{1}$ & 0.606  & 0.619 & 3032 & - & 0.600  & 0.564 & 6321 & -\\\\
$(0,1)1P(\frac{3}{2}^{-})_{1}$ & 0.612  & 0.636 & 3050 & $\Omega_{c}(3050)^{0}$/$\sim$3050/$?^{?}$~\cite{F201}   & 0.603  & 0.572 & 6329 & $\Omega_{b}(6330)^{-}$/$\sim$6330/$?^{?}$~\cite{F201}\\
$(0,1)1P(\frac{3}{2}^{-})_{2}$ & 0.615  & 0.644 & 3052 & - & 0.608  & 0.585 & 6333 & $\Omega_{b}(6340)^{-}$/$\sim$6340/$?^{?}$~\cite{F201}\\
$\mathbf{(1,0)}1P(\frac{1}{2}^{-})_{1}$ & 0.789  & 0.444 & 3055 & $\Omega_{c}(3065)^{0}$/$\sim$3065/$?^{?}$~\cite{F201} & - & - & -& - \\\\
$(0,1)1P(\frac{5}{2}^{-})_{2}$ & 0.623  & 0.670 & 3078 & $\Omega_{c}(3090)^{0}$/$\sim$3090/$?^{?}$~\cite{F201} & 0.612  & 0.596 & 6345 & $\Omega_{b}(6350)^{-}$/$\sim$6350/$?^{?}$~\cite{F201} \\\\
$\mathbf{(1,0)}1P(\frac{3}{2}^{-})_{1}$ & 0.813  & 0.485 & 3108 &$\Omega_{c}(3120)^{0}$/$\sim$3119/$?^{?}$~\cite{F201} & - & - & - & - \\\\
$(0,2)1D(\frac{1}{2}^{+})_{1}$ & 0.630  & 0.804 & 3298 & - & 0.617  & 0.726 & 6555 & - \\
$(0,2)1D(\frac{3}{2}^{+})_{2}$ & 0.634  & 0.818 & 3303 & - & 0.621  & 0.738 & 6556 & -\\
$(0,2)1D(\frac{5}{2}^{+})_{3}$ & 0.639  & 0.837 & 3307 & - & 0.625  & 0.754 & 6555 & -\\
$(0,2)1D(\frac{3}{2}^{+})_{1}$ & 0.634  & 0.818 & 3308 & - & 0.619  & 0.732 & 6561 & -\\
$(0,2)1D(\frac{5}{2}^{+})_{2}$ & 0.638  & 0.833 & 3313 & - & 0.623  & 0.745 & 6561 & -\\
$(0,2)1D(\frac{7}{2}^{+})_{3}$ & 0.643  & 0.853 & 3318 & $\Omega_{c}(3327)^{0}$/$\sim$3327/$?^{?}$~\cite{F206} & 0.628  & 0.762 & 6562 & -\\
\end{tabular}
\end{ruledtabular}
\end{table*}

\section{Conclusions}\label{sec4}

For precisely predicting the fine structure of the excitation spectra in the singly heavy baryons $\Sigma_{Q}$, $\Xi '_{Q}$ and $\Omega_{Q}$, the method adopted in this work includes the following elements: (1) Hamiltonian of the relativized quark model; (2) Wave function with the principle of HQET; (3) Specific Jacobi coordinates; (4) Improved calculation of the spin-orbit interactions taking into account of the contribution from the light-quark cluster; (5) GEM and ISG methods; (6) HQD mechanism of the orbital excitation. We systematically calculated the singly heavy baryon excitation spectra with a uniform set of parameters, and gave a consistent explanation of all the observed singly heavy baryons including the $1P$-wave excited baryons of $\Sigma_{Q}$, $\Xi '_{Q}$ and $\Omega_{Q}$.

The obtained results are as follows:
(1) The contribution of the light-quark cluster is important to configure the fine structure. (2) The HQD mechanism remains valid as a whole for singly heavy baryons. (3) $c$ quark is indeed not heavy enough, so that the HQD mechanism is broken in the $1P$-wave excited states of the $\Sigma_{c}$, $\Xi '_{c}$ and $\Omega_{c}$ baryons. (4) The fine structure of the excitation spectra for the $\Sigma_{Q}$, $\Xi '_{Q}$ and $\Omega_{Q}$ baryons are revealed. (5) The assignments of the observed baryons are performed systematically, they are $\Sigma_{c}(2800)^{++,+,0}$, $\Sigma_{c}(2846)^{0}$, $\Sigma_{b}(6097)^{+,-}$, $\Xi_{c}(2923)^{0}$, $\Xi_{c}(2939)^{0}$, $\Xi_{c}(2964)^{0}$, $\Xi_{c}(2930)^{+}$, $\Xi_{b}(6227)^{0,-}$, $\Omega_{c}(3050)^{0}$, $\Omega_{c}(3065)^{0}$, $\Omega_{c}(3090)^{0}$, $\Omega_{c}(3120)^{0}$, $\Omega_{b}(6316)^{-}$, $\Omega_{b}(6330)^{-}$, $\Omega_{b}(6340)^{-}$ and $\Omega_{b}(6350)^{-}$. (6) $\Omega_{c}(3000)^{0}$ is most likely to be interpreted as the $(0,1)1P(\frac{1}{2}^{-})_{0}$ state, even though their mass difference is relatively large. (7) $\Omega_{c}(3327)^{0}$ belongs to the $1D$-wave states, but is difficult to be assigned accurately due to the quasi-degeneracy of the energy levels.

In conclusion, with the improved calculation of spin-orbit interactions, the universality of the HQD mechanism is retested and the precise fine structures of the singly heavy excited baryons are presented. The calculation results match well with the experimental data, which implies that the method used in this work is reliable and the theoretical prediction is credible. Additionally, the contribution from the light-quark cluster to the spin-orbit interaction plays an important role in successfully describing the fine structure. We believe that the method used in this work must have already seized the most relevant degrees of freedom of the interactions.

These results are also compared with other theoretical results.
The analysis of fine structures is expected to provide a dependable theoretical reference for relevant researches. The retest of the HQD mechanism may help to understand the nature of the heavy quarks and strong interactions. The method used in this work is instructive and applicable for the study of more complex exotic hadron states, such as the heavy tetraquarks and pentaquarks.

\begin{large}
\section*{Acknowledgements}
\end{large}

 This research was supported by the Open Project of Guangxi Key Lab of Nuclear Physics and Technology (No. NLK2023-04), the Central Government Guidance Funds for Local Scientific and Technological Development in China (No. Guike ZY22096024), the Natural Science Foundation of Guizhou Province-ZK[2024](General Project)650, the National Natural Science Foundation of China (Grant Nos. 11675265, 12175068), the Continuous Basic Scientific Research Project (Grant No. WDJC-2019-13) and the Leading Innovation Project (Grant No. LC 192209000701). We thank the reviewer for valuable advice and suggestions. JG would like to thank Professor Yinsheng Ling from Soochow University for helpful discussion.


\begin{thebibliography}{}

\bibitem{F101}
F.~Gross, E.~Klempt, S.~J.~Brodsky, A.~J.~Buras, V.~D.~Burkert et al., 50 Years of Quantum Chromodynamics,
\href{https://doi.org/10.1140/epjc/s10052-023-11949-2}{Eur.Phys.J.C \textbf{83}, 1125 (2023)}, arXiv:2212.11107 [hep-ph].

\bibitem{F201}S. Navas et al., (Particle Data Group), Review of particle physics, \href{https://doi.org/10.1103/PhysRevD.110.030001}{Phys. Rev. D \textbf{110} 3 , 030001,(2024)}; R. L. Workman et al., (Particle Data Group), Review of Particle Physics,
\href{https://doi.org/10.1093/ptep/ptac097}{Prog. Theor. Exp. Phys. \textbf{2022}, 083C01,(2022) and 2023 update};

\bibitem{F210}R. Mizuk et al., (Belle Collaboration), Observation of an isotriplet of excited charmed baryons decaying to $\Lambda_{c}^{+}\pi$,
\href{https://doi.org/10.1103/PhysRevLett.94.122002}{Phys. Rev. Lett. \textbf{94}, 122002 (2005)}, arXiv:hep-ex/0412069 [hep-ex].

\bibitem{F205}B. Aubert et al., (BABAR Collaboration), A study of Excited Charm-Strange Baryons with Evidence for new Baryons $\Xi_{c}(3055)^{+}$ and $\Xi_{c}(3123)^{+}$,
\href{https://doi.org/10.1103/PhysRevD.77.012002}{Phys. Rev. D \textbf{77}, 012002 (2008)}, arXiv:0710.5763 [hep-ex].

\bibitem{F209}B. Aubert et al., (BABAR  Collaboration), Measurements of $\mathcal{B}(\bar{B}^{0}\rightarrow\Lambda_{c}^{+}\bar{p})$ and $\mathcal{B}(\bar{B}^{-}\rightarrow\Lambda_{c}^{+}\bar{p}\pi^{-})$ and Studies of $\Lambda_{c}^{+}\pi^{-}$ Resonances,
\href{https://doi.org/10.1103/PhysRevD.78.112003}{Phys. Rev. D \textbf{78}, 112003 (2008)}, arXiv:0807.4974 [hep-ex].

\bibitem{F4401}R. Aaij et al.,(LHCb Collaboration), Observation of five new narrow $\Omega_{c}^{0}$ states decaying to $\Xi_{c}^{+}K^{-}$,
\href{https://doi.org/10.1103/PhysRevLett.118.182001}{Phys. Rev. Lett. \textbf{118} 18 , 182001 (2017)}, arXiv:1703.04639 [hep-ex].

\bibitem{F4402}J. Yelton et al.,(Belle Collaboration), Observation of Excited $\Omega_{c}^{0}$ Charmed Baryons in $e^{+}e^{-}$ Collisions,
\href{https://doi.org/10.1103/PhysRevD.97.051102}{Phys. Rev. D \textbf{97} 5 , 051102 (2018)}, arXiv:1711.07927 [hep-ex].

\bibitem{F208}Y. B. Li et al., (Belle  Collaboration), Observation of $\Xi_{c}(2930)^{0}$ and updated measurement of $B^{-}\rightarrow K^{-}\Lambda_{c}^{+}\bar{\Lambda}_{c}^{-}$ at Belle,
\href{https://doi.org/10.1140/epjc/s10052-018-5720-5}{Eur. Phys. J. C \textbf{78} 3, 252 (2018)}, arXiv:1712.03612 [hep-ex].

\bibitem{F212}R. Aaij et al.,(LHCb Collaboration), Observation of Observation of a new $\Xi_{b}^{-}$ resonance,
\href{https://doi.org/10.1103/PhysRevLett.121.072002}{Phys. Rev. Lett. \textbf{121} 7, 072002 (2018)}, arXiv:1805.09418 [hep-ex].

\bibitem{F204}Y. B. Li et al., (Belle  Collaboration), Evidence of a structure in $\bar{K}^{0}\Lambda_{c}^{+}$ consistant with a charged $\Xi_{c}(2930)^{+}$, and updated measurement of $\bar{B}^{0}\rightarrow \bar{K}^{0}\Lambda_{c}^{+}\bar{\Lambda}_{c}^{-}$ at Belle,
\href{https://doi.org/10.1140/epjc/s10052-018-6425-5}{Eur. Phys. J. C \textbf{78}, 928 (2018)}, arXiv:1806.09182 [hep-ex].

\bibitem{F211}R. Aaij et al.,(LHCb Collaboration), Observation of two resonances in the $\Lambda_{b}^{0}\pi^{\pm}$ systems and precise
measurement of $\Sigma_{b}^{\pm}$ and $\Sigma_{b}^{*\pm}$ properties,
\href{https://doi.org/10.1103/PhysRevLett.122.012001}{Phys. Rev. Lett. \textbf{122} 1, 012001 (2019)}, arXiv:1809.07752 [hep-ex].

\bibitem{F4403}R. Aaij et al.,(LHCb Collaboration), First observation of excited $\Omega_{b}^{-}$ states,
\href{https://doi.org/10.1103/PhysRevLett.124.082002}{Phys. Rev. Lett. \textbf{124} 8 , 082002 (2020)}, arXiv:2001.00851 [hep-ex].

\bibitem{F203}R. Aaij et al., (LHCb Collaboration), Observation of New $\Xi_{c}^{0}$ baryons Decaying to $\Lambda_{c}^{+}K^{-}$,
\href{https://doi.org/10.1103/PhysRevLett.124.222001}{Phys. Rev. Lett. \textbf{124}, 222001 (2020)}, arXiv:2003.13649 [hep-ex].

\bibitem{F213}R. Aaij et al.,(LHCb Collaboration), Observation of Observation of a new $\Xi_{b}^{0}$ resonance,
\href{https://doi.org/10.1103/PhysRevD.103.012004}{Phys. Rev. Lett. \textbf{103} 1, 012004 (2021)}, arXiv:2010.14485 [hep-ex].

\bibitem{F2021}R. Aaij et al., (LHCb Collaboration), Observation of excited $\Omega_{c}^{0}$ baryons in $\Omega_{b}^{-}\rightarrow \Xi_{c}^{+}K^{-}\pi^{-}$ decays,
\href{https://doi.org/10.1103/PhysRevD.104.L091102}{Phys. Rev. D \textbf{104} 9 , L091102 (2021)}, arXiv:2107.03419 [hep-ex].

\bibitem{F206}R. Aaij et al.,(LHCb Collaboration), Observation of New $\Omega_{c}^{0}$ States Decaying to the $\Xi_{c}^{+}K^{-}$ Final State,
\href{https://doi.org/10.1103/PhysRevLett.131.131902}{Phys. Rev. Lett. \textbf{131} 13, 131902 (2023)}, arXiv:2302.04733 [hep-ex].

\bibitem{F207}R. Aaij et al., (LHCb Collaboration), Observation of New Baryons in the $\Xi_{b}^{-}\pi^{+}\pi^{-}$ and $\Xi_{b}^{0}\pi^{+}\pi^{-}$ Systems,
\href{https://doi.org/10.1103/PhysRevLett.131.171901}{Phys. Rev. Lett. \textbf{131} 17, 171901 (2023)}, arXiv:2307.13399 [hep-ex].

\bibitem{F214}R. Aaij et al.,(LHCb Collaboration), First determination of the spin-parity of $\Xi_{c}(3055)^{+,0}$ baryons,
\href{https://doi.org/10.48550/arXiv.2409.05440}{arXiv:2409.05440 [hep-ex]}.

\bibitem{F403}W. Roberts and M. Pervin, Heavy baryons in a quark model,
\href{https://doi.org/10.1142/S0217751X08041219}{Int. J. Mod. Phys. A \textbf{23}, 2817 (2008)}, arXiv: 0711.2492 [nucl-th].

\bibitem{Fb05}M. Karliner and J. L. Rosner, Very narrow excited $\Omega_{c}$ baryons,
\href{https://doi.org/10.1103/PhysRevD.95.114012}{Phys. Rev. D \textbf{95} 11, 114012 (2017)}, arXiv: 1703.07774 [hep-ph].

\bibitem{Fb08}E. Ortiz-Pacheco, R. Bijker, A. Giachino, and , E. Santopinto, Heavy $\Omega_{c}$ and $\Omega_{b}$  baryons in the quark model,
\href{https://doi.org/10.1088/1742-6596/1610/1/012011}{J. Phys. Conf. Ser. \textbf{1610} 1, 012011 (2020)}, arXiv: 2004.09409 [nucl-th].

\bibitem{F401}S. Godfrey and N. Isgur, Mesons in a Relativized Quark Model with Chromodynamics,
\href{https://doi.org/10.1103/PhysRevD.32.189}{Phys. Rev. D \textbf{32}, 189-231 (1985)}.

\bibitem{F402}S. Capstick and N. Isgur, Baryons in a relativized quark model with chromodynamics,
\href{https://doi.org/10.1103/PhysRevD.34.2809}{Phys. Rev. D \textbf{34}, 2809-2835 (1986)}.

\bibitem{Fp003}X. Z. Weng, W. Z. Deng, and S. L. Zhu, Heavy baryons in the relativized quark model with chromodynamics,
\href{https://doi.org/10.1103/PhysRevD.110.056052}{Phys. Rev. D \textbf{110} 5, 056052 (2024)}, arXiv: 2405.19039 [hep-ph].

\bibitem{F3171}Z. Shah, K. Thakkar, A. K. Rai, and P. C. Vinodkumar, Excited State Mass spectra of Singly Charmed Baryons,
\href{https://doi.org/10.1140/epja/i2016-16313-9}{Eur. Phys. J. A \textbf{52}, 313 (2016)}, arXiv:1602.06384 [hep-ph].

\bibitem{F3311}A. Kakadiya, Z. Shah, and A. K. Rai, Mass spectra and decay properties of singly heavy bottom-strange baryons,
\href{https://doi.org/10.1142/S0217751X22500531}{Int. J. Mod. Phys. A \textbf{37} 11n12, 2250053 (2022)}, arXiv: 2202.12048 [hep-ph].

\bibitem{F322}G. Yang and J. L. Ping, Dynamical study of $\Omega_{c}^{0}$ in the chiral quark model,
\href{https://doi.org/10.1103/PhysRevD.97.034023}{Phys. Rev. D \textbf{97} 3, 034023 (2018)}, arXiv:1703.08845 [hep-ph].

\bibitem{F323}H. X. Huang, J. L. Ping, and F. Wang, Investigating the excited $\Omega_{c}^{0}$ states through $\Xi_{c}K$ and $\Xi_{c}^{'}K$ decay channels,
\href{https://doi.org/10.1103/PhysRevD.97.034027}{Phys. Rev. D \textbf{97} 3, 034027 (2018)}, arXiv:1704.01421 [hep-ph].

\bibitem{Fp005}H. Garcilazo, J. Vijande, and  A. Valcarce, Faddeev study of heavy baryon spectroscopy,
\href{https://doi.org/10.1088/0954-3899/34/5/014}{J. Phys. G \textbf{34}, 961-976 (2007)}, arXiv: hep-ph/0703257 [hep-ph].

\bibitem{F3221}C. S. An and H. Chen, Observed $\Omega_{c}^{0}$ resonances as pentaquark states,
\href{https://doi.org/10.1103/PhysRevD.96.034012}{Phys. Rev. D \textbf{96} 3, 034012 (2017)}, arXiv:1705.08571 [hep-ph].

\bibitem{F3241}K. L. Wang,  L. Y. Xiao, X. H. Zhong, and Q. Zhao, Understanding the newly observed $\Omega_{c}$ states through their decays,
\href{https://doi.org/10.1103/PhysRevD.95.116010}{Phys. Rev. D \textbf{95} 11, 016010 (2017)}, arXiv:1703.09130 [hep-ph].

\bibitem{F324}K. L. Wang, Q. F. L\"{u}, and X. H. Zhong, Interpretation of the newly observed $\Sigma_{b}(6097)^{\pm}$ and $\Xi_{b}(6227)^{-}$ states as the $P$-wave bottom baryons,
\href{https://doi.org/10.1103/PhysRevD.99.014011}{Phys. Rev. D \textbf{99} 1, 014011 (2019)}, arXiv:1810.02205 [hep-ph].

\bibitem{Fb003}E. Santopinto, A. Giachino, J. Ferretti, H. Garc\'{\i}a-Tecocoatzi, M. A. Bedolla, R. Bijker, and E. Ortiz-Pacheco, The $\Omega_{c}$-puzzle solved by means of spectrum and decay width predictions,
\href{https://doi.org/10.1140/epjc/s10052-019-7527-4}{Eur. Phys. J. C \textbf{79} 12, 1012 (2019)}, arXiv:1811.01799 [hep-ph].

\bibitem{F334}E. Ortiz-Pacheco and R. Bijker, Masses and radiative decay widths of $S$- and $P$- wave singly, doubly, and triply heavy charm and bottom baryons,
\href{https://doi.org/10.1103/PhysRevD.108.054014}{Phys. Rev. D \textbf{108} 5, 054014 (2023)}, arXiv: 2307.04939 [hep-ph].

\bibitem{F315}J. R. Zhang, $S$-wave $D^{(*)N}$ molecular states: $\Sigma_{c}(2800)$ and $\Lambda_{c}(2940)^{+}$ ?,
\href{https://doi.org/10.1103/PhysRevD.89.096006}{Phys. Rev. D \textbf{89} 9, 096006 (2014)}, arXiv:1212.5325 [hep-ph].

\bibitem{Fb002}Z. G. Wang, Analysis of $\Omega_{c}(3000)$, $\Omega_{c}(3050)$, $\Omega_{c}(3066)$, $\Omega_{c}(3090)$ and $\Omega_{c}(3119)$ with QCD sum rules,
\href{https://doi.org/10.1140/epjc/s10052-017-4895-5}{Eur. Phys. J. C \textbf{77} 5, 325 (2017)}, arXiv:1704.01854 [hep-ph].

\bibitem{Fb06}S. S. Agaev, K. Azizi, and H. Sundu, Interpretation of the new $\Omega_{c}^{0}$ states via their mass and width,
\href{https://doi.org/10.1140/epjc/s10052-017-4953-z}{Eur. Phys. J. C \textbf{77} 6, 395 (2017)}, arXiv:1704.04928 [hep-ph].

\bibitem{Fb0021}Z. G. Wang, Analysis of $\Omega_{b}(6316)$, $\Omega_{b}(6330)$, $\Omega_{b}(6340)$ and $\Omega_{b}(6350)$ with QCD sum rules,
\href{https://doi.org/10.1142/S0217751X20500438}{Int. J. Mod. Phys. A \textbf{35} 07, 2050043 (2020)}, arXiv:2001.02961 [hep-ph].

\bibitem{Fb061}H. M. Yang and H. X Chen, $P$-wave charmed baryons of the $SU(3)$ flavor $6_{F}$,
\href{https://doi.org/10.1103/PhysRevD.104.034037}{Phys. Rev. D \textbf{104} 3, 034037 (2021)}, arXiv:2106.15488 [hep-ph].

\bibitem{F320}Q. Mao, H. X. Chen, A. Hosaka, X. Liu, and S. L. Zhu, $D$-wave heavy baryons of the $SU(3)$ flavor $\mathbf{6}_{F}$,
\href{https://doi.org/10.1103/PhysRevD.96.074021}{Phys. Rev. D \textbf{96} 7, 074021 (2017)}, arXiv:1707.03712 [hep-ph].

\bibitem{F333}Q. Xin, Z. G. Wang, and F. L\"{u}, The $\Lambda$-type $P$-wave bottom baryon states via the QCD sum rules,
\href{https://doi.org/10.1088/1674-1137/ace81f}{Chin. Phys. C \textbf{47}, 093106 (2023)}, arXiv: 2306.05626 [hep-ph].

\bibitem{F405}D. Ebert, R. N. Faustov, and V. O. Galkin, Spectroscopy and Regge trajectories of heavy baryons in the relativistic quark-diquark picture,
\href{https://doi.org/10.1103/PhysRevD.84.014025}{Phys. Rev. D \textbf{84}, 014025 (2011)}, arXiv: 1105.0583 [hep-ph].

\bibitem{F407}B. Chen, K. W. Wei, X. Liu, and T. Matsuki, Low-lying charmed and charmed-strange baryon states,
\href{https://doi.org/10.1140/epjc/s10052-017-4708-x}{Eur. Phys. J. C \textbf{77}, 154 (2017)}, arXiv: 1609.07967 [hep-ph].

\bibitem{Fp006}B. Chen and X. Liu, Assigning the newly reported $\Sigma_{b}(6097)$ as a $P$-wave excited state and predicting its partners,
\href{https://doi.org/10.1103/PhysRevD.98.074032}{Phys. Rev. D \textbf{98}, 074032 (2018)}, arXiv: 1810.00389 [hep-ph].

\bibitem{F4071}B. Chen, S. Q. Luo, and X. Liu, Universal behavior of mass gaps existing in the single heavy baryon family,
\href{https://doi.org/10.1140/epjc/s10052-021-09234-1}{Eur. Phys. J. C \textbf{81} 5, 474 (2021)}, arXiv: 2101.10806 [hep-ph].

\bibitem{F303}E. Eichten and B. R. Hill, An Effective Field Theory for the Calculation of Matrix Elements Involving Heavy Quarks,
\href{https://doi.org/10.1016/0370-2693(90)92049-O}{Phys. Lett. B \textbf{234}, 511-516 (1990)}.

\bibitem{F304}H. Georgi, An Effective Field Theory for Heavy Quarks at Low-energies,
\href{https://doi.org/10.1016/0370-2693(90)91128-X}{Phys. Lett. B \textbf{240}, 447-450 (1990)}.

\bibitem{F3231}G. Monta\~{n}a, A. Feijoo, and \`{A}. Ramos, A meson-baryon molecular interpretation for some $\Omega_{c}$ excited states,
\href{https://doi.org/10.1140/epja/i2018-12498-1}{Eur. Phys. J. A \textbf{54} 4, 64 (2018)}, arXiv:1709.08737 [hep-ph].

\bibitem{Fb04}M. Padmanath and N. Mathur, Quantum Numbers of Recently Discovered $\Omega_{c}^{0}$  Baryons from Lattice QCD,
\href{https://doi.org/10.1103/PhysRevLett.119.042001}{Phys. Rev. Lett. \textbf{119} 4 , 042001 (2017)}, arXiv:1704.00259 [hep-ph].

\bibitem{Fb01}Z. Zhao, D. D. Ye, and A. Zhang, Hadronic decay properties of newly observed $\Omega_{c}$ baryons,
\href{https://doi.org/10.1103/PhysRevD.95.114024}{Phys. Rev. D \textbf{95} 11 , 114024 (2017)}, arXiv:1704.02688 [hep-ph].

\bibitem{Fb011}Q. F. L\"{u}, Canonical interpretations of the newly observed $\Xi_{c}(2923)^{0}$, $\Xi_{c}(2939)^{0}$, and $\Xi_{c}(2965)^{0}$ resonances,
\href{https://doi.org/10.1140/epjc/s10052-020-08488-5}{Eur. Phys. J. C \textbf{80} 10, 921 (2020)}, arXiv:2004.02374 [hep-ph].


\bibitem{Fp007}R. Bijker,  H. Garc\'{\i}a-Tecocoatzi, A. Giachino,  E. Ortiz-Pacheco, and E. Santopinto, Masses and decay widths of $\Xi_{_{c/b}}$ and $\Xi_{c/b}^{'}$ baryons,
\href{https://doi.org/10.1103/PhysRevD.105.074029}{Phys. Rev. D \textbf{105}, 074029 (2022)}, arXiv: 2010.12437 [hep-ph].

\bibitem{Fb07}H. Garc\'{\i}a-Tecocoatzi, A. Giachino, A. Ramirez-Morales, A. Rivero-Acosta, and E. Santopinto, Decay widths and mass spectra of single bottom baryons,
\href{https://doi.org/10.48550/arXiv.2307.00505}{ arXiv:2307.00505 [hep-ph]}.

\bibitem{F410}B. Chen, S. Q. Luo, X. Liu, and T. Matsuki, Interpretation of the observed $\Lambda_{b}(6146)^{0}$ and $\Lambda_{b}(6152)^{0}$ states as $1D$ bottom baryons,
\href{https://doi.org/10.1103/PhysRevD.100.094032}{Phys. Rev. D \textbf{100}, 094032 (2019)}, arXiv: 1910.03318 [hep-ph].

\bibitem{F502}G. L. Yu, Z. Y. Li, Z. G. Wang, J. Lu, and M. Yan, Systematic analysis of single heavy baryons $\Lambda_{Q}$, $\Sigma_{Q}$ and $\Omega_{Q}$,
\href{https://doi.org/10.1016/j.nuclphysb.2023.116183}{Nucl. Phys. B \textbf{990}, 116183 (2023)}, arXiv: 2206.08128 [hep-ph].

\bibitem{F503}Z. Y. Li, G. L. Yu, Z. G. Wang, J. Z. Gu, J. Lu, and H. T. Shen, Systematic analysis of strange single heavy baryons $\Xi_{c}$ and $\Xi_{b}$,
\href{https://doi.org/10.1088/1674-1137/acd365}{Chin. Phys. C \textbf{47}, 073105 (2023)}, arXiv: 2207.04167 [hep-ph].

\bibitem{F504}G. L. Yu, Z. Y. Li, Z. G. Wang, J. Lu, and M. Yan, Systematic analysis of doubly charmed baryons $\Xi_{cc}$ and $\Omega_{cc}$,
\href{https://doi.org/10.1140/epja/s10050-023-01044-1}{Eur. Phys. J. A \textbf{59}, 126 (2023)}, arXiv: 2211.00510 [hep-ph].

\bibitem{F505}Z. Y. Li, G. L. Yu, Z. G. Wang, J. Z. Gu, and H. T. Shen, Mass spectra of double-bottom baryons,
\href{https://doi.org/10.1142/S0217732323500529}{Mod. Phys. Lett. A \textbf{38} 08n09,2350052 (2023)}, arXiv: 2210.13085 [hep-ph].

\bibitem{F506}Z. Y. Li, G. L. Yu, Z. G. Wang, J. Z. Gu, and H. T. Shen, Mass spectra of bottom-charm baryons,
\href{https://doi.org/10.1142/S0217751X23500951}{Int. J. Mod. Phys. A \textbf{38} 18n19, 2350095 (2023)}, arXiv: 2211.15111 [hep-ph].

\bibitem{F501}Z. Y. Li, G. L. Yu, Z. G. Wang, and J. Z. Gu, Heavy quark dominance in orbital excitation of singly and doubly heavy baryons,
\href{https://doi.org/10.1140/epjc/s10052-024-12457-7}{Eur. Phys. J. C \textbf{84} 2, 106 (2024)}, arXiv: 2311.08251 [hep-ph].

\bibitem{F906}M. G. Mayer, On Closed Shells in Nuclei. II, Phys. Rev. \textbf{75}, 1969 (1949); O. Haxel, J. H. D. Jensen, and H. E. Suess, On the ``Magic Numbers'' in Nuclear Structure, Phys. Rev. \textbf{75}, 1766 (1949).

\bibitem{F905}S. Acharya et al. (The ALICE Collaboration), Evidence of Spin-Orbital Angular Momentum interactions in Relativistic Heavy-Ion Collisions,
\href{https://doi.org/10.1103/PhysRevLett.125.012301}{Phys. Rev. Lett. \textbf{125} 1, 012301(2020)}, arXiv:1910.14408 [nucl-ex].

\bibitem{F904}V. F. Maisi, A. Hofmann, M. R\"{o}\"{o}sli, J. Basset, C. Reichl, W. Wegscheider, T. Ihn, and K. Ensslin, Spin-Orbit Coupling at the Level of a Single Electron,
\href{https://doi.org/10.1103/PhysRevLett.116.136803}{Phys. Rev. Lett. \textbf{116}, 136803 (2016)}, arXiv:1512.05149 [cond-mat.mes-hall]; J. M. Due\~{n}as, J. H. Garc\'{\i}a, and S. Roche, Emerging Spin-Orbit Torques in Low-Dimensional Dirac Materials, \href{https://doi.org/10.1103/PhysRevLett.132.266301}{Phys. Rev. Lett. \textbf{132}, 266301 (2024)}, arXiv:2310.06447 [cond-mat.mes-hall].

\bibitem{F903}J. B. Curtis, N. R. Poniatowski, Y. L. Xie, A. Yacoby, E. Demler, and P. Narang, Stabilizing Fluctuating Spin-Triplet Superconductivity in Graphene via Induced Spin-Orbit Coupling,
\href{https://doi.org/10.1103/PhysRevLett.130.196001}{Phys. Rev. Lett. \textbf{130}, 196001 (2023)}, arXiv:2209.10560 [cond-mat.supr-con]; M. Amundsen, J. Linder, J. W. A. Robinson, I. \v{Z}uti\'{c}, and N. Banerjee, Spin-orbit effects in superconducting hybrid structures, \href{https://doi.org/10.1103/RevModPhys.96.021003}{Rev. Mod. Phys. \textbf{96} 2, 021003(2024)}, arXiv:2210.03549 [cond-mat.supr-con].

\bibitem{F902}S. Takahashi, Y. Kubo, K. Konishi, R. Issaoui, J. Barjon, and Nobuko Naka, Spin-Orbit Effects on Exciton Complexes in Diamond,
\href{https://doi.org/10.1103/PhysRevLett.132.096902}{Phys. Rev. Lett. \textbf{132}, 096902 (2024)}.

\bibitem{Fp001}N. Isgur and G. Karl, P-Wave Baryons in the Quark Model,
\href{https://doi.org/10.1103/PhysRevD.18.4187}{Phys. Rev. D \textbf{18}, 4187 (1978)}.

\bibitem{Fp002}N. Isgur, Meson-like baryons and the spin-orbit puzzle,
\href{https://doi.org/10.1103/PhysRevD.62.014025}{Phys. Rev. D \textbf{62}, 014025 (2000)}, arXiv: hep-ph/9910272 [hep-ph].

\bibitem{F306}S. Capstick and W. Roberts, Quark models of baryon masses and decays,
\href{https://doi.org/10.48550/arXiv.nucl-th/0008028}{Prog. Part. Nucl. Phys. \textbf{45}, S241-S331 (2000)}, arXiv:nucl-th/0008028 [nucl-th].

\bibitem{F601}Q. F. L\"{u}, D. Y. Chen, and Y. B. Dong, Masses of doubly heavy tetraquarks $T_{QQ'}$ in a relativized quark model,
\href{https://doi.org/10.1103/PhysRevD.102.034012}{Phys. Rev. D \textbf{102}, 034012 (2020)}, arXiv: 2006.08087 [hep-ph].

\bibitem{F6021}M. Kamimura, Nonadiabatic coupled-rearrangement-channel approach to muonic molecules,
\href{https://doi.org/10.1103/PhysRevA.38.621}{Phys. Rev. A \textbf{38}, 621-624 (1988)}.

\bibitem{F602}E. Hiyama, Y. Kino, and M. Kamimura, Gaussian expansion method for few-body systems,
\href{https://doi.org/10.1016/S0146-6410(03)90015-9}{Prog. Part. Nucl. Phys. \textbf{51}, 223-307 (2003)}.

\bibitem{F603}T. Yoshida, E. Hiyama, A. Hosaka, M. Oka, and  K. Sadato, Spectrum of heavy baryons in the quark model,
\href{https://doi.org/10.1103/PhysRevD.92.114029}{Phys. Rev. D \textbf{92}, 114029 (2015)}, arXiv: 1510.01067 [hep-ph].

\bibitem{F604}C. Q. Pang,  J. Z. Wang, X. Liu, and  T. Matsuki, A systematic study of mass spectra and strong decay of strange mesons,
\href{https://doi.org/10.1140/epjc/s10052-017-5434-0}{Eur. Phys. J. C \textbf{77} 12, 861 (2017)}, arXiv: 1705.03144 [hep-ph].


\bibitem{F327}E. Braaten, L. P. He, and A. Mohapatra, Masses of doubly heavy tetraquarks with error bars,
\href{https://doi.org/10.1103/PhysRevD.103.016001}{Phys. Rev. D \textbf{103}, 016001 (2021)}, arXiv:2006.08650 [hep-ph].

\bibitem{Fp004}N. Isgur and G. Karl, Symmetry Breaking in Baryons,
\href{https://doi.org/10.1016/0370-2693(78)90676-7}{Phys. Lett. B \textbf{74}, 353-356 (1978)}.

\bibitem{F411}H. Y. Cheng, Charmed baryon physics circa 2021,
\href{https://doi.org/10.1016/j.cjph.2022.06.021}{Chin. J. Phys. \textbf{78}, 324-362 (2022)}, arXiv:2109.01216 [hep-ph].

\end{thebibliography}
\end{document}